\documentclass[useAMS,usenatbib]{mn2e}
\usepackage{graphicx,amssymb,amstext} 
\usepackage{hyperref}

\DeclareRobustCommand{\ion}[2]{%
\relax\ifmmode
\ifx\testbx\f@series
{\mathbf{#1\,\mathsc{#2}}}\else
{\mathrm{#1\,\mathsc{#2}}}\fi
\else\textup{#1\,{\mdseries\textsc{#2}}}%
\fi}

\def\aj{AJ}			
\def\araa{ARA\&A}		
\def\apj{ApJ}		
\def\apjl{ApJ}		
\def\apjs{ApJS}				
		
\def\aap{A\&A}		
		
\def\aaps{A\&AS}

\def\mnras{MNRAS}

\def\pasp{PASP}

\begin{document}

\title[Dependence of SFR indicators on galaxy physical parameters]{{\it GAMA/H-ATLAS}: Common star-formation rate indicators and their dependence on galaxy physical parameters}

\author[L.~Wang et al.]
{\parbox{\textwidth}{\raggedright L.~Wang$^{1,2,3}$\thanks{E-mail: \texttt{l.wang@sron.nl}},
P. Norberg$^{3}$, M. L. P. Gunawardhana$^{3}$, S. Heinis$^{4}$,  
I. K. Baldry$^{5}$, 
J. Bland-Hawthorn$^{6}$, 
N. Bourne$^{7}$,
S. Brough$^{8}$, 
M. J. I. Brown$^{9}$,  
M. E. Cluver$^{10}$,
A. Cooray$^{11}$, 
E. da Cunha$^{12}$, 
S. P. Driver$^{13, 14}$,
 L. Dunne$^{7, 15, 16}$,
 S. Dye$^{17}$, 
S. Eales$^{16}$, 
M. W. Grootes$^{18}$,
B. W. Holwerda$^{19}$,
A. M. Hopkins$^{8}$,
E. Ibar$^{20}$,
R. Ivison$^{7, 21}$, 
C. Lacey$^{3}$,
M. A. Lara-Lopez$^{22}$,  
J. Loveday$^{23}$, 
S. J. Maddox${^{7, 15, 16}}$,
M. J. Micha{\l}owski$^{7}$,
I. Oteo$^{7, 21}$,
M. S. Owers$^{8, 24}$,
C. C. Popescu$^{25}$, 
D. J. B. Smith$^{26}$, 
E. N. Taylor$^{27}$, 
R. J. Tuffs$^{18}$,     
P. van der Werf$^{28}$}\vspace{0.4cm}\\
\parbox{\textwidth}{\raggedright $^{1}$SRON Netherlands Institute for Space Research, Landleven 12, 9747 AD, Groningen, The Netherlands\\
$^{2}$Kapteyn Astronomical Institute, University of Groningen, Postbus 800, 9700 AV Groningen, the Netherlands\\
$^{3}$Institute for Computational Cosmology, Department of Physics, Durham University, Durham, DH1 3LE, UK\\
$^{4}$Department of Astronomy, University of Maryland, College Park, MD 20742-2421, USA\\
$^{5}$Astrophysics Research Institute, Liverpool John Moores University, IC2, Liverpool Science Park, 146 Brownlow Hill, Liverpool, L3 5RF, UK\\
$^{6}$Sydney Institute for Astronomy, School of Physics A28, University of Sydney, NSW 2006, Australia\\
$^{7}$Institute for Astronomy, University of Edinburgh, Royal Observatory, Edinburgh EH9 3HJ, UK\\
$^{8}$Australian Astronomical Observatory, PO Box 915, North Ryde, NSW 1670, Australia\\
$^{9}$School of Physics and Astronomy, Monash University, Clayton, Victoria 3800, Australia\\
$^{10}$The University of Western Cape, Robert Sobukwe Road, Bellville 7530, South Africa\\
$^{11}$Department of Physics \& Astronomy, University of California, Irvine, CA 92697, USA\\
$^{12}$Centre for Astrophysics and Supercomputing, Swinburne University of Technology, Hawthorn, Victoria 3122, Australia\\
$^{13}$International Centre for Radio Astronomy Research (ICRAR), University of Western Australia, Crawley, WA 6009, Australia\\ 
$^{14}$SUPA, School of Physics \& Astronomy, University of St Andrews, North Haugh, St Andrews, KY16 9SS, UK\\
$^{15}$Department of Physics and Astronomy, University of Canterbury, Private Bag 4800, Christchurch, New Zealand\\
$^{16}$School of Physics and Astronomy, Cardiff University, The Parade, Cardiff CF24 3AA, UK\\
$^{17}$School of Physics and Astronomy, University of Nottingham, University Park, Nottingham, NG7 2RD, UK\\
$^{18}$Max Planck Institut fuer Kernphysik, Saupfercheckweg 1, 69117 Heidelberg, Germany\\
$^{19}$University of Leiden, Sterrenwacht Leiden, Niels Bohrweg 2, NL-2333 CA Leiden, The Netherlands\\
$^{20}$Instituto de F\'isica y Astronom\'ia, Universidad de Valpara\'iso, Avda. Gran Breta\~na 1111, Valpara\'iso, Chile\\
$^{21}$European Southern Observatory, Karl Schwarzschild Strasse 2, D-85748 Garching, Germany\\
$^{22}$Instituto de Astronom'a, Universidad Nacional Aut\'onoma de M\'exico, A.P. 70-264, 04510 M\'exico, D.F., M\'exico\\
$^{23}$Astronomy Centre, University of Sussex, Falmer, Brighton BN1 9QH, UK\\
$^{24}$Department of Physics and Astronomy, Macquarie University, NSW 2109, Australia\\
$^{25}$Jeremiah Horrocks Institute, University of Central Lancashire, PR1 2HE, UK\\
$^{26}$Centre for Astrophysics, Science \& Technology Research Institute, University of Hertfordshire, Hatfield, Herts, AL10 9AB\\
$^{27}$School of Physics, The University of Melbourne, Victoria 3010, Australia\\
$^{28}$Leiden Observatory, Leiden University, P.O. Box 9513, NL-2300 RA Leiden, The Netherlands}}

\date{Accepted . Received ; in original form }

\maketitle

\pagebreak
\newpage
\newpage

\clearpage

\begin{abstract}

\vspace*{-0.2cm}

We compare common star-formation rate (SFR) indicators in the local Universe in the GAMA equatorial fields ($\sim160$ deg$^2$), using  ultraviolet (UV) photometry from GALEX,  far-infrared (FIR) and sub-millimetre (sub-mm) photometry from H-ATLAS, and H$\alpha$ spectroscopy from the GAMA  survey. With a high-quality sample of 745 galaxies  (median redshift $\left<z\right>=0.08$), we consider three SFR tracers: UV luminosity corrected for dust attenuation using the UV spectral slope $\beta$ (SFR$_{\rm UV, corr}$), H$\alpha$  line luminosity corrected for dust using the Balmer decrement (BD) (SFR$_{\rm H\alpha, corr}$), and the combination of UV and IR emission (SFR$_{\rm UV + IR}$).  We demonstrate that SFR$_{\rm UV, corr}$ can be reconciled with the other two tracers after applying attenuation corrections by calibrating IRX (i.e. the IR to UV luminosity ratio)  and attenuation in the H$\alpha$ (derived from BD) against $\beta$. However, $\beta$ on its own is very unlikely to be a reliable attenuation indicator. We find that attenuation correction factors depend on parameters such as stellar mass ($M_*$), $z$ and dust temperature ($T_{\rm dust}$), but not on H$\alpha$ equivalent width (EW) or Sersic index. Due to the large scatter in the IRX vs $\beta$ correlation, when compared to SFR$_{\rm UV + IR}$, the $\beta$-corrected SFR$_{\rm UV, corr}$ exhibits systematic deviations as a function of IRX, BD and $T_{\rm dust}$.

\end{abstract}

\begin{keywords}
galaxies: photometry - galaxies: statistics - infrared: galaxies - ultraviolet: galaxies- methods: statistical 
\end{keywords}

\section{INTRODUCTION}

The distribution functions of star-formation rates (SFR) at different cosmic epochs (or more commonly its integrated form, the evolution of the cosmic star-formation rate density) provide fundamental observational tests for theoretical models of galaxy formation and evolution (e.g., Hopkins \& Beacom 2006; Madau \& Dickinson 2014). Observationally,  one can use a variety of indicators at different wavelengths to measure the level of star-formation activity in galaxies. The rest-frame ultraviolet  (UV) non-ionising stellar continuum luminosity, where newly formed massive stars emit the bulk of their energy, is often used as a direct SFR indicator, especially at high redshift. H$\alpha$ nebular recombination emission line luminosity, which probes the hydrogen-ionising photons produced by the most massive and short-lived  stars, is another commonly used SFR indicator when spectroscopy is available.

One of the most significant challenges when using UV or H$\alpha$ line emission as a direct star-formation tracer is the effect of dust attenuation as the process of star formation takes place in dense, cold and often dusty molecular gas clouds.  Indeed, some of the most intensely star-forming galaxies are extremely UV-faint, e.g., those selected in the sub-millimetre (sub-mm). To overcome this problem, various empirical or semi-empirical correction methods have been developed to determine the amount of dust attenuation in a galaxy. For example, the power-law spectral slope of the rest-frame UV continuum $\beta$ ($f^{\lambda}\propto \lambda^{\beta}$), or its proxy the FUV - NUV colour, has been widely used as a practical method for estimating the global attenuation corrections (e.g., Meurer et al. 1995, 1997, 1999; Burgarella et al. 2005;  Laird et al. 2005; Reddy et al. 2006; Salim et al. 2007; Treyer et al. 2007; Wijesinghe et al. 2011). However, the spectral slope $\beta$ shows a wide dispersion with varying dust properties, dust/star geometry and redshift (e.g., Witt \& Gordon 2000; Granato et al. 2000; Oteo et al. 2014). In addition, $\beta$ is sensitive to the intrinsic UV spectral slope (determined by properties such as age of the stellar populations, star-formation history, metallicity, etc.) and as such is dependent on a number of parameters that are not solely related to dust attenuation (e.g., Kong et al. 2004; Buat et al. 2005). 

Another independent method to correct for dust is to use Balmer decrement ratio measurement (i.e. the observed flux ratio of the H$\alpha$ and H$\beta$ nebular emission lines) to estimate the amount of dust attenuation at H$\alpha$ (e.g., Kennicutt 1992; Brinchmann et al. 2004; Moustakas et al. 2006; Garn \& Best 2010). However, Balmer decrement measurements are generally only available for bright H II regions within the galaxies, and so can be problematic when applying to the whole galaxy. Also, H$\beta$ is considerably weaker than H$\alpha$. The Balmer decrement method is also found to be a poor estimator for dust attenuation in dusty starbursts (e.g., Moustakas, Kennicutt \& Tremonti 2006).

From an energy conservation point of view, one can also derive SFR from dust emission as dust absorbs the UV and optical light from newly formed stars and re-emit predominantly in the far-infrared (FIR) and sub-mm. The main advantage of inferring SFR from the IR emission is that it is not affected by dust attenuation. However, some of the IR emission could be caused by heating from the old/evolved stellar populations or AGN and thus is not related to recent star formation (e.g., Helou 1986; Popescu et al. 2000; Bell et al. 2003; Natale et al. 2015). 

{The aim of this paper is to take advantage of the Galaxy and Mass Assembly (GAMA)\footnote{\url{http://www.gama-survey.org}} survey (Driver et al. 2009, 2011; Liske et al. 2015) and associated multi-wavelength surveys to carefully examine some of the most commonly used SFR indicators. GAMA provides a large sample of galaxies in the local Universe where photometric information in the UV and IR as well as measurements of key spectral lines such as H$\alpha$ and H$\beta$ are available. The paper is organised as follows. In Section 2, we describe the various surveys and derived data products used in our analysis. In Section 3, we give a brief overview of the three SFR indicators investigated in this paper and our selection criteria in the UV bands, optical emission lines, and IR and sub-mm bands. In Section 4, we study in detail the properties of our galaxy samples selected at different wavelengths and construct a joint UV-H$\alpha$-IR sample. Then focusing on the joint sample, we examine the dust attenuations derived using different methods, and the correlations between various SFR indicators as a function of galaxy physical parameters such as stellar mass, redshift, UV continuum slope, Balmer decrement, IRX (i.e. the total IR to UV luminosity ratio), S\'ersic index, H$\alpha$ emission line equivalent width, and dust temperature. Finally, we give conclusions and discussions in Section 5. In an upcoming GAMA paper (Luke et al., in prep.), 12 SFR metrics are examined and calibrated to a mean relation. In this paper, we focus on just three commonly used SFR indicators and inter-compare them using a very high quality galaxy sample. With random statistical errors minimised, we investigate the influence of different systematic errors on these SFR indicators.

Throughout the paper, we assume a flat $\Lambda$CDM cosmological model with $\Omega_M=0.3$,  $\Omega_{\Lambda}=0.7$,  $H_0=70$ km s$^{-1}$ Mpc$^{-1}$. Flux densities are corrected for Galactic extinction using the $E(B-V)$ values provided by Schlegel, Finkbeiner \& Davis (1998). We use the AB magnitude scale and the Kroupa (Kroupa \& Weidner 2003) initial mass function (IMF) unless otherwise stated. 

\section{Sample selection}

\subsection{Spectroscopic \& multi-wavelength photometric data}

GAMA is an optical spectroscopic survey of low-redshift galaxies, mainly conducted at the Anglo-Australian Telescope. GAMA covers three equal-sized fields to an apparent SDSS DR7 (Sloan Digital Sky Survey - Data Release 7) Petrosian r-band magnitude limit of $r=19.8$ mag:  G09, G12 and G15 (centred at a right ascension of $\sim$9, 12, and 14.5 hours, respectively), on the celestial equator. We include all GAMA II main survey targets (SURVEY\_CLASS $>=4$) with reliable AUTOZ  redshifts ($nQ>=3$; Baldry et al. 2014) from the tiling catalogue version 45 (Baldry et al. 2010).  We impose a lower redshift limit of 0.01 which is used to remove stars and galaxies for which peculiar motion would overwhelm the Hubble flow, leading to highly uncertain distances based upon recession velocities alone. We also impose an upper redshift limit of 0.5 as there are very few GAMA galaxies above this redshift. After applying these cuts and restricting galaxies to areas with both GALEX and {\it Herschel} coverage, we are left with a total of 128,170 galaxies which forms our parent sample in this paper.

In addition to the spectroscopic survey, GAMA has also assembled imaging data from a number of independent surveys in order to generate multi-wavelength photometric information spanning the wavelength range from 1 nm to 1 mm. Below we summarise the two main imaging surveys that are relevant to this paper.

\subsubsection{GALEX}

The Galaxy Explorer Mission (GALEX, Martin et al. 2005) conducted a number of major surveys and observer motivated programs, most notably the all-sky imaging survey (AIS) and the medium imaging survey (MIS). A dedicated programme (the GALEX guest investigator programme GALEX-GAMA) provided further GALEX observations in GAMA fields to MIS depth (1500s). The final collated data provides near-complete NUV and FUV coverage of the primary GAMA II regions.  In the three equatorial regions, coverage is at the 90\% level in both FUV and NUV bands. The GAMA-GALEX ultraviolet catalogue is a combination of archival data and pointed observations on equatorial GAMA fields. The archival data have been used to extend the ultraviolet coverage of the GAMA regions as much as possible. In this paper, we use the GAMA GALEX catalogue described in Liske et al. (2015).The full width at half maximum (FWHM) of the point spread function (PSF) is $\sim$ 4.2 and 5.3 arcsec in the FUV and NUV, respectively. As the resolution of the GALEX observations is significantly lower than that of the SDSS data, a variety of methods are employed to address the source blending issue and derive the FUV and NUV fluxes for every GAMA galaxy. For more details, we refer the reader to Liske et al. (2015). 

\subsubsection{H-ATLAS}

The {\it Herschel} (Pilbratt et al. 2010) Astrophysical Terahertz Large Area Survey (H-ATLAS) survey (Eales et al. 2010) conducted observations of the three equatorial fields also observed in the  GAMA redshift survey. H-ATLAS images were obtained using {\it Herschel}'s fast-scan parallel mode, where the spacecraft is moving at 60" per sec, with Photometric Array Camera and Spectrometer (PACS; Poglitsch et al. 2010) observing simultaneously at 100 and 160 $\mu$m and  Spectral and Photometric Imaging Receiver (SPIRE; Griffin et al. 2010) at 250, 350 and 500 $\mu$m.  The FWHM of the PSF of the telescope  is 9, 13, 18, 25 and 35 arc sec at 100, 160, 250, 350 and 500 $\mu$m, respectively (Ibar et al. 2010; Griffin et al. 2010).  The total area covered by H-ATLAS in the GAMA 09, 12 and 15 fields is approximately 160 deg$^2$. Details of the map-making process can be found in Valiante et al. (in prep).

The PACS and SPIRE photometry are derived from the PACS/SPIRE maps by measuring the flux in the appropriate optically defined aperture convolved with the appropriate PACS/SPIRE PSF. Where objects overlap care is taken to divide the flux between the two objects following the prescription outlined in Appendix A of Bourne et al. (2012) and Driver et al. (2015). The process also involves measuring flux through a range of randomly located apertures for a range of aperture sizes, and fitting a simple parametric function (2nd order polynomial) to the median flux as a function of aperture area. This contaminating flux level is then subtracted from the target flux to provide a background corrected flux. The final error in these bands is then half of 1-$\sigma$ quantile range (from 16\% to 84\%) of the flux as a function of aperture. This implicitly assumes that the dominant error is the subtraction of contaminating flux (see Driver et al. 2016 for more details). 

In addition to the forced sub-mm photometry at known optical source positions, in Appendix~\ref{appendix1}, we also consider the blind H-ATLAS source catalogue matched with GAMA galaxies through the likelihood ratio (LR) technique. We show that the use of the LR-matched GAMA/H-ATLAS catalogue does not change our conclusions.

\subsection{Properties of GAMA galaxies}

The GAMA team has produced a number of catalogues of galaxy physical properties derived from the amassed multi-wavelength photometric and spectroscopic datasets. Below we summarise the properties of the two catalogues which we make use of in this paper.

The stellar masses ($M_{\rm star}$) in GAMA (Taylor et al. 2011) are estimated based on SED fitting of aperture (AUTO) photometry in the the rest-frame wavelength range between 3000 and 11000  \AA \ (approximately $u$ through $Y$) using a grid of synthetic spectra. The spectra are generated by assuming exponentially decaying star-formation history, the Bruzual \& Charlot (2003) stellar evolution models with the Chabrier (2003) IMF and the Calzetti et al. (2000) dust extinction law. In order to be consistent with SFR calculations in this paper which assumes the Kroupa IMF, we apply a correction to the stellar masses using $M^{\rm Kroupa}_{\rm star} = M^{\rm Chabrier}_{\rm star}  / 0.94$. 

A single-Sersic (1-component only) fit is performed across all passbands in the SDSS ($ugriz$), the UKIRT Infrared Deep Sky Survey Large Area Survey (UKIDSS-LAS; Lawrence et al. 2007) ($YJHK$) and VISTA Kilo-degree Infrared Galaxy survey (VIKING) ($ZYJHK$) surveys. Galaxy single-Sersic modelling is achieved using SIGMA v1.0-2 (Structural Investigation of Galaxies via Model Analysis) (Kelvin et al. 2012). SIGMA is a wrapper around several contemporary astronomy tools such as Source Extractor (Bertin \& Arnouts 1996), PSF Extractor (Bertin \& Delorme, priv. comm.) and GALFIT 3 (Peng et al. 2010). GALFIT is the workhorse fitting algorithm used within SIGMA. It uses a simple downhill minimisation in order to minimise the global $\chi^2$ of the model.

\section{Dust corrected SFR indicators}

There are many methods to derive the SFR of a galaxy. Most of them rely on simple linear conversions from luminosities at a given wavelength. In this paper we consider three SFR indicators widely used in the literature, i.e., UV continuum luminosity corrected for dust attenuation using the UV spectral slope (SFR$_{\rm UV, corr}$),  H$\alpha$ emission line luminosity corrected for dust attenuation using the Balmer decrement (SFR$_{\rm H\alpha, corr}$), and combination of the escaped and dust re-processed emission by adding together UV and infrared continuum luminosity (SFR$_{\rm UV + IR}$). We examine the overall statistical correlations between SFR$_{\rm UV + IR}$ and SFR$_{\rm UV, corr}$ as well as between SFR$_{\rm UV + IR}$ and SFR$_{\rm H\alpha, corr}$. Then, we investigate the ratios between different SFR indicators as a function of various galaxy physical parameters. 

\subsection{SFR from UV continuum emission}

\begin{figure}
\includegraphics[height=2.5in,width=3.6in]{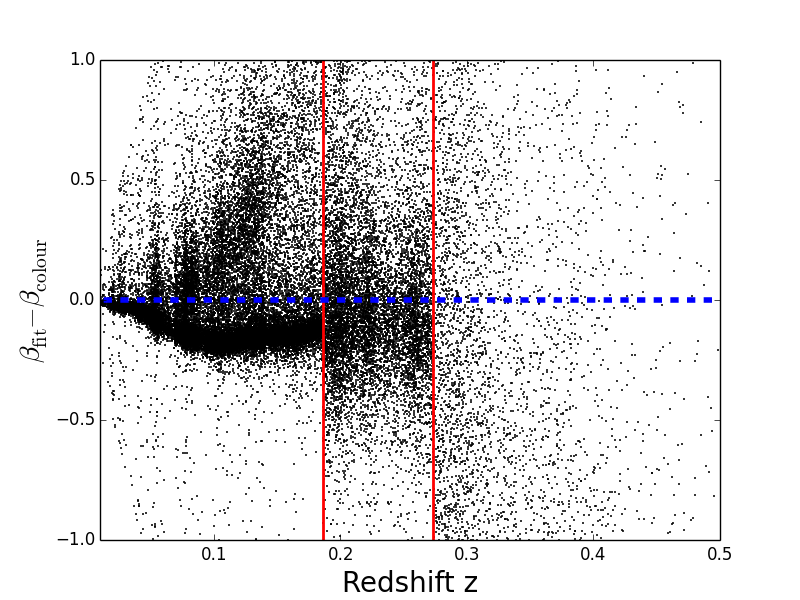}
\caption{The difference in the $\beta$ estimates ($\Delta \beta = \beta_{\rm fit} - \beta_{\rm colour}$) as a function of redshift for galaxies with S/N $\geq 5$ in both  NUV and FUV and at least two broad-band filters in the rest-frame UV with which to derive $\beta_{\rm fit}$. The blue dashed line corresponds to $\beta_{\rm fit} = \beta_{\rm colour}$.  The red vertical lines mark $z=0.185$ (where the SDSS $u$-band shifts into the rest-frame UV) and $z=0.273$ (where the FUV band shifts out of the rest-frame UV and so the only common band between $\beta_{\rm fit}$ and $\beta_{\rm colour}$ is the NUV band).} 
\label{fig:beta_comparison_z}
\end{figure}

\begin{figure}
\includegraphics[height=2.5in,width=3.45in]{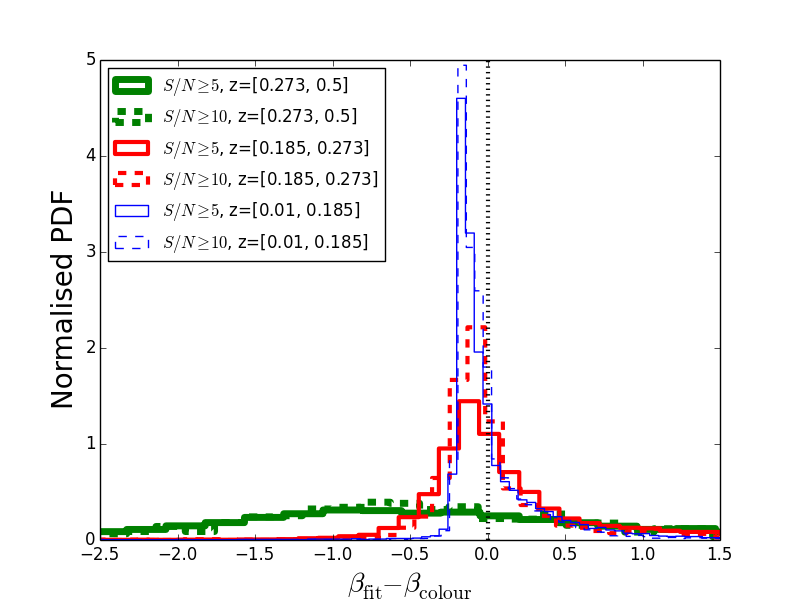}
\caption{The normalised histogram of $\Delta\beta=\beta_{\rm fit} - \beta_{\rm colour}$ in three redshift bins (blue: $0.01<z<0.185$; red $0.185<z<0.273$;  green: $0.273<z<0.5$) for galaxies with at least two filters in the rest-frame UV. The solid histograms correspond to S/N $\geq5$ in FUV and NUV. The dashed histograms correspond to S/N $\geq10$ in FUV and NUV. The vertical dotted line is where $\beta_{\rm colour}=\beta_{\rm fit}$. It is clear that above $z=0.273$, there is a large systematic difference as well as a large scatter in $\Delta\beta$.}
\label{fig:beta_comparison}
\end{figure}

\begin{table}
\caption{The 16th, 50th and 84th percentile in $\Delta \beta=\beta_{\rm fit} - \beta_{\rm colour}$ in three redshift bins. We also examine the impact of  S/N in the FUV and NUV on $\Delta \beta$.}\label{table:beta}
\begin{tabular}{lll}
\hline
 & $S/N\geq5$& $S/N\geq10$\\
\hline
$z=[0.01, 0.185] $     &  -0.17, -0.09, 0.29 &  -0.17, -0.08, 0.18 \\
\hline
$z=[0.185, 0.273] $    & -0.28, -0.01, 0.72  & -0.23, -0.06, 0.38 \\
\hline
$z=[0.273, 0.5] $       & -1.85, -0.55, 0.94  &   -1.78, -0.60, 0.64    \\
\hline
\end{tabular}
\end{table}

To estimate the dust attenuation correction in the FUV using the observed FUV-NUV colour, we use the empirical relation in \citet{hao11}
\begin{equation}\label{eq_afuv}
A_{\rm FUV} = 3.83 \times [({\rm FUV} - {\rm NUV})_{\rm obs} -0.022].
\end{equation}
This relation is based on calibrating  the total IR to FUV luminosity ratio (IRX) and the attenuation in H$\alpha$ line luminosity (derived from Balmer decrement) against the FUV - NUV colour using a nearby normal star-forming galaxy sample (Moustakas \& Kennicutt 2006), which was designed to cover the full range of optical spectral characteristics present in the local galaxy population. For a detailed discussion on the range of applicability of this relation and comparison with other $A_{\rm FUV}$--IRX relations in the literature (e.g., Meurer et al. 1999; Kong et al. 2004; Buat et al. 2005; Burgarella et al. 2005), please refer to Hao et al. (2011).
To estimate dust attenuation in the NUV, $A_{\rm NUV}$, we use the \citet{Calzetti_2000} attenuation law, which yields a ratio $A_{\rm  FUV}/A_{\rm NUV} = 1.245$. While \citet{hao11} parameterised the dust attenuation using broad-band GALEX colour, we will also consider an equivalent parametrisation later using the power-law slope of the UV continuum, $\beta$, which we derive using two methods.

The first method relies on the colour relation from \citet{Kong_2004}: 
\begin{equation}\label{eq_beta}
\beta = (\log f^{\lambda}_{\rm FUV} - \log f^{\lambda}_{\rm NUV}) / (\log \lambda_{\rm FUV} - \log \lambda_{\rm NUV}) 
\end{equation}
with $f^{\lambda}_{\rm FUV}$ and $f^{\lambda}_{\rm NUV}$ referring to the monochromatic flux (or flux density) per unit wavelength interval, and effective wavelengths $\lambda_{\rm FUV} =1528$ \AA ~and $\lambda_{\rm NUV} =2271$  \AA.  If using flux densities per unit frequency interval, then $\beta$ can be derived as
\begin{equation}\label{eq_beta_v2}
\beta = (\log f^{\nu}_{\rm FUV} - \log f^{\nu}_{\rm NUV}) / (\log \lambda_{\rm FUV} - \log \lambda_{\rm NUV})  - 2.
\end{equation}
We refer hereafter to this estimate derived using Eq. 2 or Eq. 3 as $\beta_{\rm colour}$. We can replace $({\rm FUV} - {\rm NUV})_{\rm obs}$ with $\beta$ in Eq. \ref{eq_afuv} using Eq. \ref{eq_beta} or Eq. \ref{eq_beta_v2}:
\begin{equation}\label{eq_afuv_beta_fit}
A_{\rm FUV} = 1.65 \times \beta + 3.22.
\end{equation}
As our galaxy sample spans a considerable range in redshift, it is necessary to apply $K$-corrections to the observed FUV-NUV colour, $({\rm FUV} - {\rm NUV})_{\rm obs}$.  $K$-corrections as derived from Loveday et al. (2012) using KCORRECT v4\_2 (Blanton \& Roweis 2007) are applied in the GALEX bands. We also consider another estimate of $\beta$ by directly fitting the available photometric data with $f^{\lambda} \propto \lambda ^{\beta}$ (or equivalently $f^{\nu} \propto \lambda ^{\beta+2}$) in the rest-frame UV range ($1200<\lambda<3000$ \AA). This enables us to use all the available data, and the relevant constraints on $\beta$ at a given redshift. For instance, at $z\gtrsim 0.185$, the SDSS u band (3557 \AA) will be used in the power-law fit and above $z=0.273$, the GALEX FUV band will move out of our fitting window. We refer hereafter to this estimate as $\beta_{\rm fit}$. In this paper, we prefer $\beta_{\rm fit}$ over $\beta_{\rm colour}$ as no $K$-corrections are needed in the derivation of $\beta_{\rm fit}$. As a result, $\beta_{\rm fit}$ should not be affected by the uncertainties in the $K$-corrections in the UV bands.

Fig.~\ref{fig:beta_comparison_z} shows the difference in the two $\beta$ estimates, $\Delta \beta =\beta_{\rm fit} - \beta_{\rm colour}$, as a function of redshift. We only show galaxies with signal-to-noise (S/N) $\geq 5$ in both the NUV and FUV bands and at least two broad-band filters in the rest-frame UV with which to derive $\beta_{\rm fit}$.  Below $z\sim0.185$, the two estimates are similar to each other but there is a small systematic difference. As the same photometric information is used in deriving $\beta_{\rm fit}$ and $\beta_{\rm colour}$ at $z<0.185$, the small systematic difference is only due to the the difference between the UV spectral shape used in KCORRECT and the power-law approximation used in deriving $\beta_{\rm fit}$. At $z\gtrsim0.185$, the $u$-band shifts into the rest-frame UV causing further scatter around the systematic difference seen at $z<0.185$. At $z\gtrsim0.273$, the FUV band shifts out of the rest-frame UV and so the NUV band is the only common filter used in both types of $\beta$ estimates, which clearly causes even larger scatter and a large systematic difference between $\beta_{\rm colour}$ and $\beta_{\rm fit}$. 

To see this more clearly, Fig.~\ref{fig:beta_comparison} shows the normalised  histogram in $\Delta\beta$ in three redshift bins for galaxies with at least two broad-band filters in the rest-frame UV. We also investigate the impact of S/N in the FUV and NUV on $\Delta\beta$.  The solid histograms correspond to galaxies with S/N $\geq5$ in FUV and NUV, while the dashed histograms correspond to galaxies with S/N $\geq10$ in FUV and NUV. In the lowest two redshift bins, i.e. $z=[0.01, 0.185]$ and $z=[0.185, 0.273]$, there is a small systematic difference between $\beta_{\rm fit}$ and $\beta_{\rm colour}$. $\beta_{\rm colour}$ systematically overestimates the UV continuum slope compared to $\beta_{\rm fit}$, which means that  the inferred dust correction will be systematically higher than using $\beta_{\rm fit}$. The scatter in $\Delta\beta$ in the medium redshift bin $z=[0.185, 0.273]$ is larger than the lowest redshift bin $z=[0.01, 0.185]$ which is understandable as the $u$-band is used in deriving $\beta_{\rm fit}$ at $z>0.185$. The agreement in the two $\beta$ estimates improves for galaxies selected at higher S/N in the UV bands.  In the highest redshift bin $z=[0.273, 0.5]$, there is a much larger systematic difference as well as a much larger scatter. In Table 1, we list the 16th, 50th, and 84th percentile in $\Delta\beta$ in the three redshift bins, $z=[0.01, 0.185]$, $[0.185, 0.273]$, and $[0.273, 0.5]$.  For galaxies at $z<0.273$, the median systematic difference between the two $\beta$ estimates is small ($<0.1$). In contrast, the median systematic difference in $\Delta\beta$ is very large ($>0.5$) for galaxies at $z>0.273$. Therefore, we do not consider those galaxies above $z=0.273$ any further in this paper.

In the following sections, we define a UV-selected sample by selecting galaxies in the redshift range $z=[0.01, 0.273]$ detected in both FUV and NUV at a S/N of 10 or greater and with at least two broad-band filters in the rest-frame UV with which to derive $\beta_{\rm fit}$. This UV-selected sample contains a total of 16,920 galaxies, i.e. 19.5\% of the parent sample defined in Section 2.1 (after excluding galaxies in the parent sample at $z>0.273$). The median statistical error (due to measurement error of the flux density) on $\beta$ ($\beta_{\rm fit}$ or $\beta_{\rm colour}$) of this sample is 0.2, which is very similar to the nearby star-forming galaxy sample (Moustakas \& Kennicutt 2006) studied in Hao et al. (2011). In Section 4.1, we define a joint sample by applying the UV, H$\alpha$, and IR selection criteria together. The median statistical error on $\beta$ ($\beta_{\rm fit}$ or $\beta_{\rm colour}$) for the joint sample is considerably smaller, 0.1. However, we point out that the statistical error on $\beta_{\rm colour}$ is likely to be underestimated for our high S/N sample, as we have ignored the uncertainties introduced in applying $K$-corrections in the UV bands.\footnote{Some studies (e.g., Rosario et al. 2016) do not apply $K$-corrections in the UV bands as they are expected to be small ($<30\%$ out to $z=0.15$). We note that for our high S/N sample, $K$-corrections actually dominate over pure measurement error. However, as $K$-corrections in the UV are very uncertain, it is not clear which approach is better.}

With GAMA spectroscopic redshifts, we can derive both FUV and NUV luminosities, as $\nu L_{\nu}$, from the GALEX photometry and correct them for dust attenuation,
\begin{equation}
L_{\rm UV, corr}  = 10^{0.4\times A_{\rm UV}} L_{\rm UV}, 
\end{equation}
where UV is either FUV or NUV.
Using these dust corrected luminosities, we derive SFR using the calibrations in Hao et al. (2011), Murphy et al. (2011) and summarised Kennicutt \& Evans (2012),
\begin{equation}
{\rm SFR}_{\rm FUV, corr} / M_{\odot} {\rm yr}^{-1} =1.72 \times 10^{-10} L_{\rm FUV, corr} / L_{\odot},
\end{equation}
and
\begin{equation}
{\rm SFR}_{\rm NUV, corr} / M_{\odot} {\rm yr}^{-1} =2.60 \times 10^{-10} L_{\rm NUV, corr} / L_{\odot},
\end{equation}
which assumes the Kroupa IMF. The statistical errors on the UV-based SFRs are calculated by simply propagating the statistical errors on the FUV and NUV flux and the statistical error on $\beta_{\rm colour}$ or $\beta_{\rm fit}$.

\subsection{SFR from H$\alpha$ emission line}

As outlined in Hopkins et al. (2003) and Gunawardhana et al. (2013), in order to measure SFR for the whole galaxy using the H$\alpha$ recombination line luminosity, corrections for the underlying Balmer stellar  absorption, dust obscuration as well as  the aperture sampled by the fibre are required,
\begin{eqnarray}
L_{\rm H\alpha, corr} ({\rm Watts}) & = & (EW_{H\alpha} + EW_c) \times 10^{-0.4 (M_r-34.1)} \\ \nonumber
                                      & & \times \frac{3\times10^{18}}{[6564.61(1+z)]^2} \left(\frac{F_{H\alpha}/F_{H\beta}}{2.86}\right)^{2.36},
\end{eqnarray}
where $EW_{H\alpha}$ and $EW_c$ are the positive (i.e. emission) H$\alpha$ equivalent width and the constant equivalent width correction, $M_r$ is the absolute $r$-band magnitude, $F_{H\alpha}/F_{H\beta}$ is the Balmer decrement.  Briefly, the H$\alpha$-based SFRs used in this investigation are based on $EW_{H\alpha}$. A correction for the missing flux due to the size of the fibre aperture is applied to each galaxy using their $r$--band absolute magnitudes to estimate the continuum luminosity at the wavelength of H$\alpha$. As this approach to applying aperture corrections, described in detail in Hopkins et al. (2003), relies on the assumption that the $r$--band continuum traces the distribution of H$\alpha$ emission within a galaxy, the correction can overestimate or underestimate the line luminosity. Based on a large sample of GAMA galaxies with $z<0.05$ that has also been observed with the SAMI (Sydney-AAO Multi-object Integral-field spectrograph) instrument (Croom et al. 2012; Bryant et al. 2015), Richards et al. (2016) demonstrated that the SFRs corrected for aperture effects following the method described in Eq. 8 tend to on average overestimate SFRs by $\sim0.1$ dex. This is in agreement of Brough et al. (2013) who found that the SFRs derived based on Eq. 8 can be overestimated on average by a factor of 1.26, based on observations of 18 GAMA galaxies using the SPIRAL optical integral field unit (IFU) on the Anglo-Australian Telescope. The aperture effect, of course, reduces with increasing redshift, and the median redshift of our joint  UV-H$\alpha$-IR sample defined in Section 4.1 is 0.08 (around 25\% of our joint sample is at $z<0.05$). In the paper, we do not attempt to apply a size-dependent aperture correction due to the large scatter in the correction itself. A constant correction ($EW_c=2.5$ \AA) for the underlying Balmer stellar absorption in H$\alpha$ EWs is incorporated\footnote{See Gunawardhana et al. (2013) for a discussion on the impact of the assumption of constant stellar absorption corrections on the H$\alpha$ line luminosities.}. Balmer line fluxes (e.g.\,H$\alpha$ and H$\beta$) are used in the calculation of dust obscuration corrections for the H$\alpha$ luminosities.  For galaxies with $F_{H\alpha}/F_{H\beta}< 2.86$, no attenuation correction is applied (i.e. $F_{H\alpha}/F_{H\beta}$ is set to 2.86).  Some of the galaxies in the sample have Balmer decrements less than 2.86, which can result from intrinsically low reddening combined with stellar absorption and flux calibration and line flux measurement errors (e.g., Kewley  et al. 2006). Furthermore, the theoretical case B value can be lower than 2.86 for galaxies hosting high temperature HII regions (e.g., L{\'o}pez-S{\'a}nchez \& Esteban 2009).  The H$\alpha$-based SFRs, corrected for dust attenuation using the measured Balmer decrement, is derived using the following calibration (Hao et al. 2011; Murphy et al. 2011; Kennicutt \& Evans 2012), 
\begin{equation}
{\rm SFR}_{\rm H\alpha, corr} / M_{\odot} {\rm yr}^{-1}  = 2.07 \times 10^{-8} L_{\rm H\alpha, corr} / L_{\odot},
\end{equation}
based on the Kroupa IMF.  Compared to ${\rm SFR}_{\rm H\alpha, corr}$ values published in Gunawardhana et al. (2013) which uses the Kennicutt (1998) calibration and the Salpeter IMF (Salpeter 1955), the new calibration in Kennicutt \& Evans (2012) is a factor of 0.68 lower.}

We exclude galaxies dominated by emission from AGNs from this analysis as their H$\alpha$ SFRs based on EWs can be contaminated by the AGN emission. The strong optical emission line (e.g.\,[\ion{N}{ii}] $\lambda6584$/H$\alpha$ and [\ion{O}{iii}] $\lambda5007$/H$\beta$) diagnostics (BPT; Baldwin, Phillips \& Terlevich 1981) and the theoretical AGN and star-forming/composite discrimination prescription of Kewley et al. (2001) can be used to identify AGNs. The more conservative discrimination prescription of Kauffmann et al. (2003) can be used to further identify pure star-forming galaxies from the star-forming/composite population identified with the Kewley et al. (2001) method. In the cases of galaxies where one or more of the four optical emission lines necessary for this type of diagnostic are not available, the two-line diagnostics are used to identify AGNs. The galaxies that cannot be classified as either AGN, composite or SF remain in the sample, as a galaxy with detected H$\alpha$ emission but without an [\ion{N}{ii}] $\lambda6584$ or [\ion{O}{iii}] $\lambda5007$ measurement is more likely to be star-forming than AGN (Cid Fernandes et al. 2010).

The GAMA emission-line sample spans $0<z<0.34$. Above $z=0.34$, the H$\alpha$ feature is either at the end of the spectral range (and so can be erroneous) or is redshifted out of the spectrum. In the following sections, we define an H$\alpha$-selected sample by requiring S/N $\geq15$ for both H$\alpha$ and H$\beta$ emission lines for all star-forming galaxies using the Kauffmann et al. (2003) classification. This is similar to the selection criteria used in Hao et al. (2011).  To match the redshift range of the UV-selected sample in Section 3.1, we select galaxies at $z=[0.01, 0.273]$. Additionally, we remove all galaxies in the $0.15<z<0.17$ range from the sample. Over this redshift range, the redshifted H$\alpha$ emission line overlaps with the atmospheric O$_2$ band, which in some cases can lead to an overestimated H$\alpha$ measurements. This H$\alpha$-selected sample contains a total of 5,171 galaxies, which is 6.7\% of the parent sample defined in Section 2.1 (after excluding galaxies in the parent sample at $0.15<z<0.17$ and $z>0.273$).

\subsection{SFR from adding UV and IR emission}

\begin{figure}
\includegraphics[height=2.6in,width=3.5in]{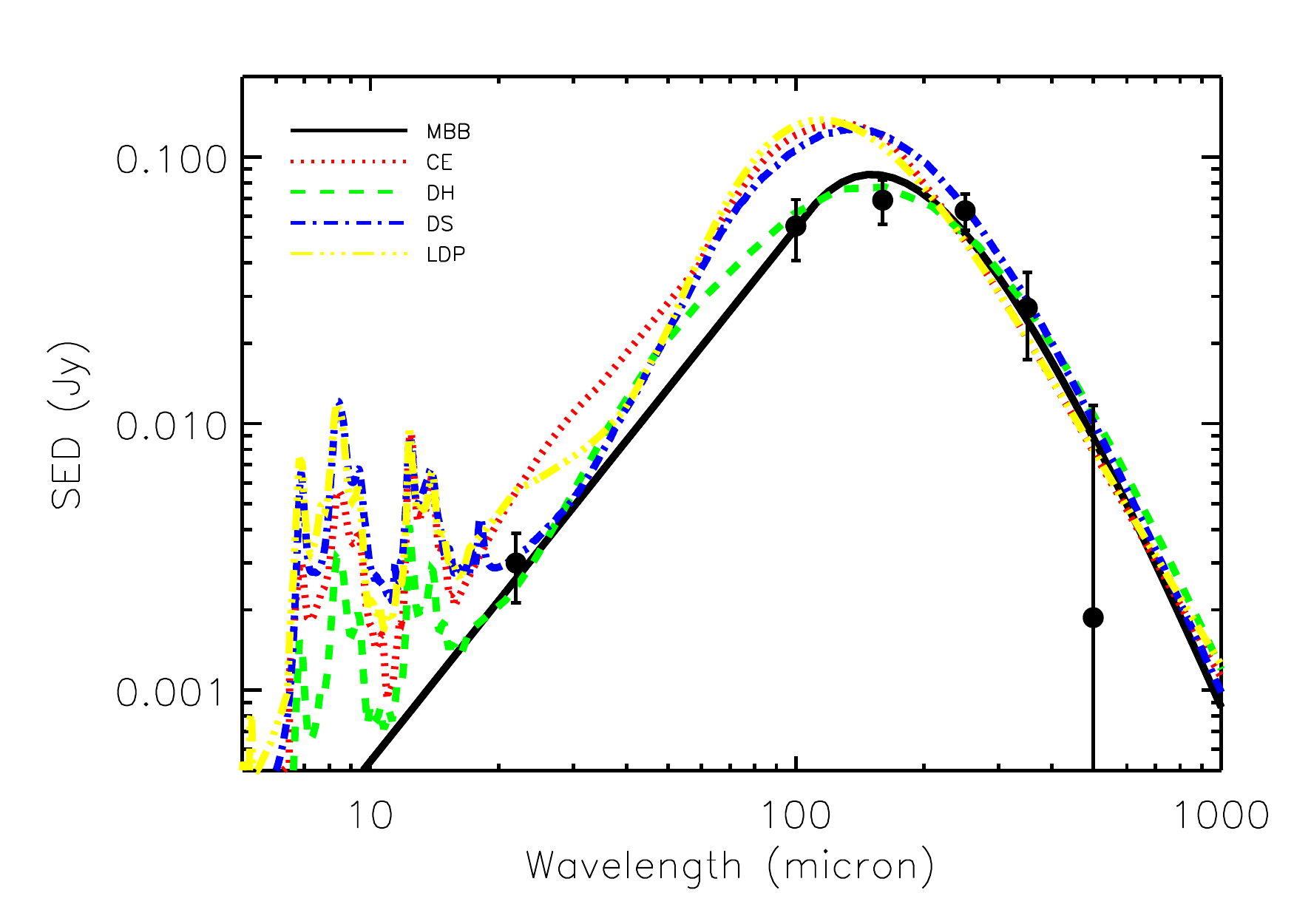}
\caption{Example SED fit  of a randomly chosen galaxy from the IR-selected sample in the observed frame. The filled dots and errors bars correspond to the measured flux densities and errors at WISE 22 $\mu$m and  {\it Herschel} wavelengths (100, 160, 250, 350 and 500 $\mu$m). The different coloured lines are the best-fit SED from different libraries.}
\label{fig:sedfit}
\end{figure}

\begin{figure}
\includegraphics[height=2.5in,width=3.45in]{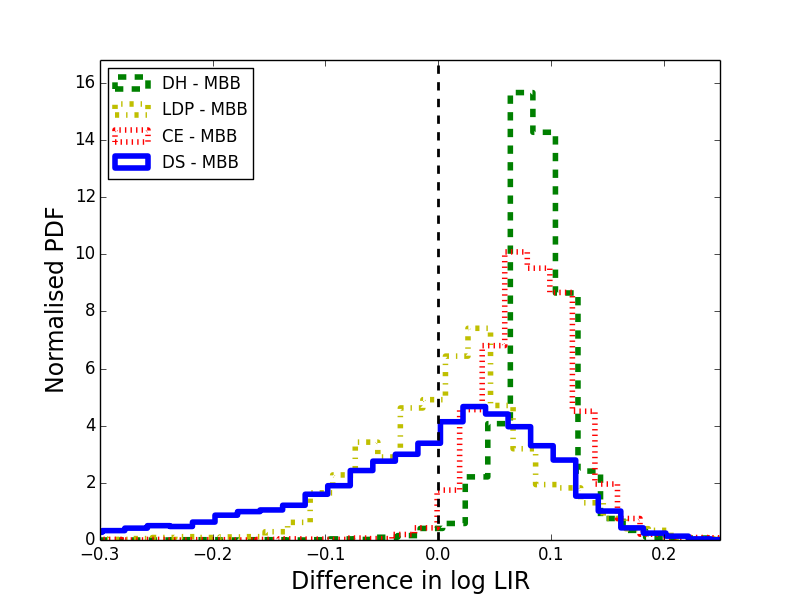}
\caption{The normalised histogram of the difference in the infrared luminosity (in unit of $L_{\odot}$) for the IR-selected sample estimated from SED fitting using different SED libraries. The dashed line is where there is no systematic difference. The systematic differences in $L_{\rm IR}$ arising from different libraries is relatively small.}
\label{fig:lir_hist}
\end{figure}

\begin{table}
\caption{The 16th, 50th and 84th percentile in the difference  of the total infrared luminosity estimates $\Delta \log L_{\rm IR}$  using different SED libraries. In the MBB library, we use a default value of $\beta=2$. We also examine $\Delta \log L_{\rm IR}$ with and without the addition of the WISE 22 $\mu$m photometry. The median difference in $\Delta \log L_{\rm IR}$ is very small in all cases ($<0.1$ dex).}\label{table:beta}
\begin{tabular}{lll}
\hline
&No  22 $\mu$m   & With 22 $\mu$m   \\
\hline
DH - MBB &0.03, 0.08, 0.12& 0.07, 0.09, 0.11\\
\hline
LDP - MBB &-0.05, -0.00, 0.06&-0.06, 0.02, 0.08\\
\hline 
CE - MBB &0.01, 0.05, 0.10&0.04, 0.08, 0.12\\
\hline
DS - MBB &-0.13, -0.04, 0.03 &-0.12 0.01, 0.09 \\
\hline
\end{tabular}
\end{table}

We can derive the total SFR of a galaxy by adding together the unobscured star formation traced by the observed UV continuum emission and the obscured star formation traced by the infrared dust emission. This method is built on an energy balance consideration which argues that all the starlight absorbed at UV and optical wavelengths by interstellar dust grains is re-emitted in the IR and sub-mm (e.g., Popescu et al. 2000; Tuffs et al. 2004; Bianchi et al. 2008; Baes et al. 2010, 2011; Holwerda et al. 2012; De Looze et al. 2012, 2014). However, there are also limitations with this method, e.g. the contribution of the old stellar populations and asymmetric star/dust geometries (e.g., Bell 2003). These effects may average out for a large statistical sample.

The infrared luminosity $L_{\rm IR}$ is defined as the integrated luminosity from rest-frame 8 to 1000 $\mu$m. To estimate the $L_{\rm IR}$ of our GAMA galaxies, we have performed SED fitting to WISE 22 $\mu$m and {\it Herschel} photometry from 100 to 500 $\mu$m. There are many suites of empirical models and templates that describe the IR SEDs of galaxies. In this paper, we consider five different SED libraries, the Chary \& Elbaz (2001; CE) templates, the Dale \& Helou (2002; DH) templates, the Lagache, Dole \& Puget  (2003; LDP) templates, the Smith et al. (2012; DS) templates and the modified blackbody (MBB) templates. The CE library contains 105 template SEDs of different luminosity classes generated to reproduce the observed correlation between mid-infrared and far-infrared luminosities (from 7 to 850 $\mu$m) for local galaxies. The DH library contains 64 locally calibrated templates for normal star-forming galaxies, differing in the slope of the power law distribution of dust mass and dust emissivity as a function of  the radiation field intensity. The LDP library contains 92 template spectra of starburst and normal star-forming galaxies, which are constructed  and optimised to reproduce statistical quantities like number counts, redshift distributions and the cosmic infrared background. The LDP templates for normal star-forming galaxies all have the same shape and are only scaled in luminosity. The DS library is based on a sample of 250 $\mu$m selected galaxies at $z<0.5$ from the H-ATLAS survey. It is worth noting that the CE and DH libraries are constructed from IRAS-selected samples which favour galaxies with a larger warm dust component. In comparison, the H-ATLAS based DS library includes star-forming galaxies that are much colder. For the MBB templates, we follow the parameterisation in Hall et al. (2010). We assume that the IR SED is a greybody at low frequencies, and a power law at high frequencies, i.e.
\begin{equation}
\Theta(\nu) = \nu^{\beta} B_{\nu}(T_d), \nu<\nu_{0}
\end{equation}
and
\begin{equation}
\Theta(\nu) = \nu^{-\gamma}, \nu\geq\nu_{0}
\end{equation}
where $\beta$ is the dust emissivity index and $B_{\nu}$ is the Planck function with an effective dust temperature $T_d$. These two functions (Eq. 10 and Eq. 11) are joined at frequencies $\nu_{0}$ that can be solved from
\begin{equation} 
\frac{d\ln{[\nu_{\beta}B_{\nu}(T_d)]}}{d\ln{\nu}} = -\gamma,
\end{equation}
with $\gamma=2$ (Hall et al. 2010). We also fix $\beta=2$ (Draine \& Lee 1984; Mathis \& Whiffen 1989)\footnote{We have also tried  lower $\beta$ values, $\beta=1.5$ and $\beta=1.2$ (e.g., Dunne \& Eales 2001; Planck Collaboration et al. 2011). We find that the difference in the resulting $L_{\rm IR}$ is very small. The mean value of $\log L_{\rm IR} (\beta=2) - \log L_{\rm IR} (\beta=1.5)$ and $\log L_{\rm IR} (\beta=2) - \log L_{\rm IR} (\beta=1.2)$ is -0.03 dex and -0.05 dex, respectively.}.

 \begin{figure}
\includegraphics[height=2.3in,width=3.2in]{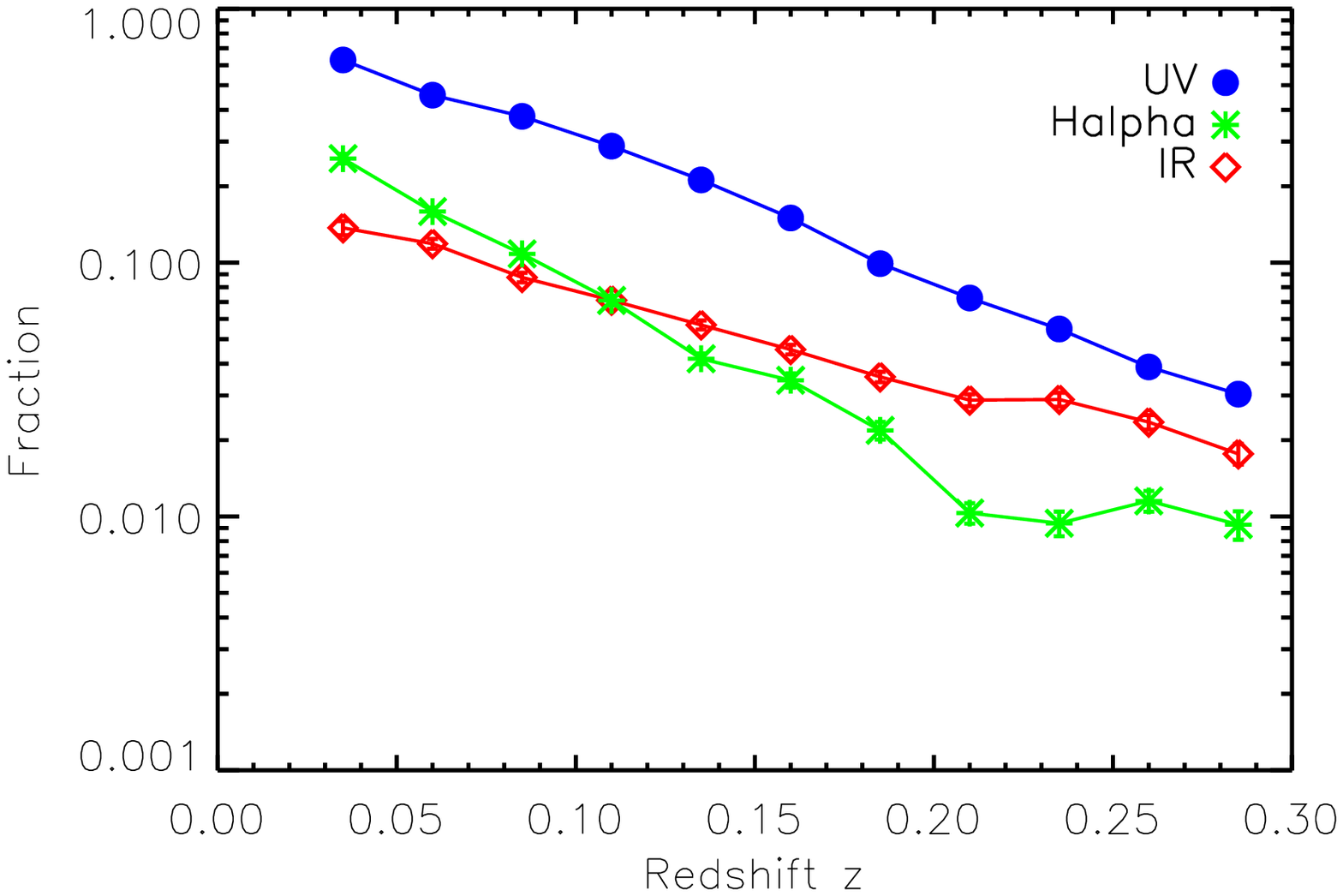}
\includegraphics[height=2.3in,width=3.2in]{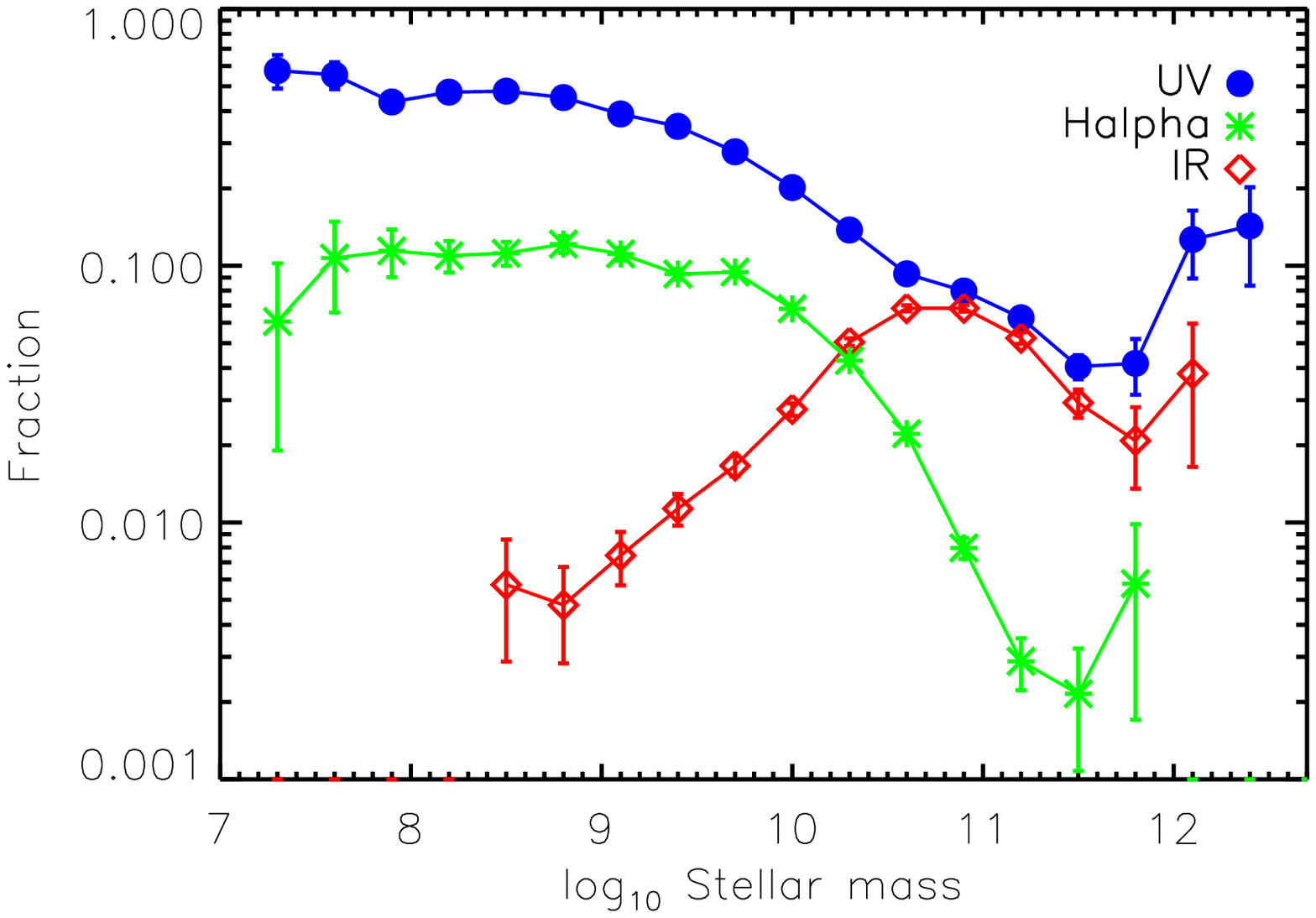}
\caption{Top: The fraction of galaxies selected at different wavelengths (blue filled dots: UV: green asterisks: H$\alpha$; red diamonds: IR) compared to the parent sample (defined in Section 2.1) as a function of redshift. Bottom: The fraction of selected galaxies compared to the parent sample as a function of stellar mass over $0.01<z<0.273$. For the H$\alpha$ selected sample, galaxies with $0.15<z<0.17$ in the parent sample are excluded.}
\label{fig:parent_vs_selected}
\end{figure}

Within each SED library, we select the best-fit template by finding the one with the lowest $\chi^2$ value, allowing for rescaling of the template. The error on the $L_{\rm IR}$ derived using the MBB templates is calculated by marginalising over the effective dust temperature $T_d$ parameter. The MBB library is the only one out of the five considered here which has a continuous parameter ($T_d$) characterising the shape of the SED. The other four libraries (i.e., CE, DH,  LDP and DS) have discrete templates which make it difficult to derive marginalised error on the $L_{\rm IR}$. For this reason and also the fact that the MBB library has a larger IR colour range than the other libraries, we choose the $L_{\rm IR}$ values derived using the MBB library as our default in the following sections. It is worth pointing out that the lack of polycyclic aromatic hydrocarbons (PAH) emission features in the MBB templates can potentially cause $L_{\rm IR}$ to be systematically underestimated. Shipley et al. (2013) estimated that the median ratio of PAH luminosity (the sum of emission features at 6.2, 7.7, 8.6, 11.3, 12.7 and 17.0 $\mu$m) to $L_{\rm IR}$ is 0.09 for a sample of star-forming IR-luminous galaxies. However, the distribution of $L_{\rm PAH}$/$L_{\rm IR}$ is quite wide. In some galaxies, $L_{\rm PAH}$/$L_{\rm IR}$ could be as high as over $20\%$ or as low as $0\%$.

To match the UV-selected sample in Section 3.1, we construct an infrared-selected sample in the redshift range between $z=[0.01, 0.273]$ by requiring galaxies detected at WISE 22 and {\it Herschel} SPIRE 250 $\mu$m at a S/N of 3 or greater. In addition, we require a {\it Herschel} PACS detection with S/N $\geq3$ at 100 or 160 $\mu$m. So, we have a good sampling of the dusty SED in the MIR, close to the peak, and in the Rayleigh-Jeans tail. This IR selected sample contains a total of 5,182 galaxies, i.e. 6.0\% of the parent sample defined in Section 2.1 (after excluding galaxies $z>0.273$). The median statistical error on $\log L_{\rm IR}$ (due to measurement error of the flux densities) for this selected sample is 0.05 dex. For the same sample, the median error on $\log L_{\rm IR}$ is 0.07 dex without the WISE 22 $\mu$m constraint. For the joint UV-H$\alpha$-IR sample defined in Section 4.1, the median statistical error on $\log L_{\rm IR}$ is 0.03 dex.

In Fig.~\ref{fig:sedfit}, we show an example SED fit of a randomly chosen galaxy from the infrared-selected sample described above. The different coloured lines correspond to the best-fit SED from the different SED libraries. Fig.~\ref{fig:lir_hist} shows the normalised histogram of the difference in $\log L_{\rm IR}$ between various libraries for the IR-selected sample. In Table 2, we list the 16th, 50th and 84th percentile in $\Delta \log L_{\rm IR}$. It is clear that the systematic differences in $L_{\rm IR}$ arising from different SED libraries is relatively small. The median value in $\Delta\log_{L_{\rm IR}}$ is $<0.1$ dex with or without adding the WISE 22 $\mu$m constraint. As mentioned above, the CE and DH libraries are biased towards galaxies which contain large warm dust content. As a result, the $L_{\rm IR}$ values derived from the CE and DH libraries are systematically higher than the values derived from the MBB library. It is also clear from Table 2 that the lack of PAH features in the MBB templates is not the main cause of the systematic difference between the CE (or the DH) library and the MBB library as the other two libraries (LDP and DS) also have PAH emission features.

Now we can add the obscured star formation traced by infrared emission and the unobscured/escaped star formation  traced by UV emission to form an estimate of the total SFR of a galaxy. Using the calibrations in Hao et al. (2011) and Kennicutt \& Evans (2012), we have
\begin{eqnarray}
{\rm SFR}_{\rm FUV+IR} / M_{\odot} {\rm yr}^{-1} & = &1.72 \times 10^{-10}  [L_{\rm FUV} / L_{\odot} \nonumber \\
&  & +\ 0.46  \times L_{\rm IR} / L_{\odot}], 
\end{eqnarray}
and
\begin{eqnarray}
{\rm SFR}_{\rm NUV+IR} / M_{\odot} {\rm yr}^{-1} & = & 2.60 \times 10^{-10}  [L_{\rm NUV} / L_{\odot} \nonumber \\
&  & +\ 0.27  \times L_{\rm IR} / L_{\odot}], 
\end{eqnarray}
based on the Kroupa IMF. We emphasise that the $L_{\rm FUV}$ and $L_{\rm NUV}$ in Eq. 13 and 14 are the observed luminosity not corrected for dust attenuation.

\begin{table*}
\caption{The sample selection criteria in different wavebands. Applying all selection criteria in the UV, H$\alpha$ and IR leaves us a total sample of 745 objects.}\label{table:selection}
\begin{tabular}{lll}
\hline
Wavelength & Selection Criteria   &  Number of galaxies   \\
\hline
UV & $z=[0.01, 0.273]$, S/N $\geq10$ in FUV \&NUV, at least 2 broad-band filters in the rest-frame UV&  16,920\\
\hline
H$\alpha$& $z=[0.01, 0.15]$ or $[0.17, 0.273]$, S/N $\geq15$ in H$\alpha$ \& H$\beta$, star-forming galaxies &  5,171\\
\hline
IR &$z=[0.01, 0.273]$, S/N $\geq3$ at 250 and 22 $\mu$m, and S/N $\geq3$ at either 100 or 160 $\mu$m&  5,182\\
\hline
\end{tabular}
\end{table*}

\begin{figure}
\includegraphics[height=2.4in,width=3.4in]{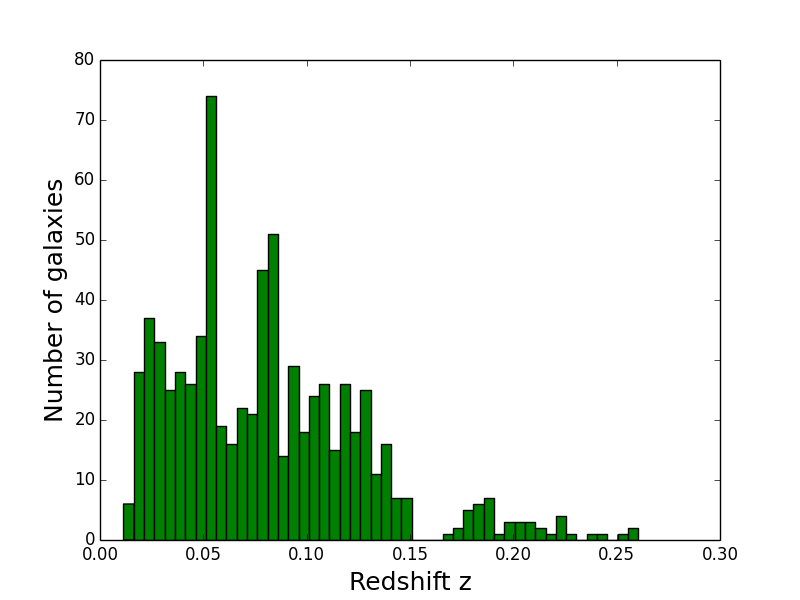}
\includegraphics[height=2.4in,width=3.4in]{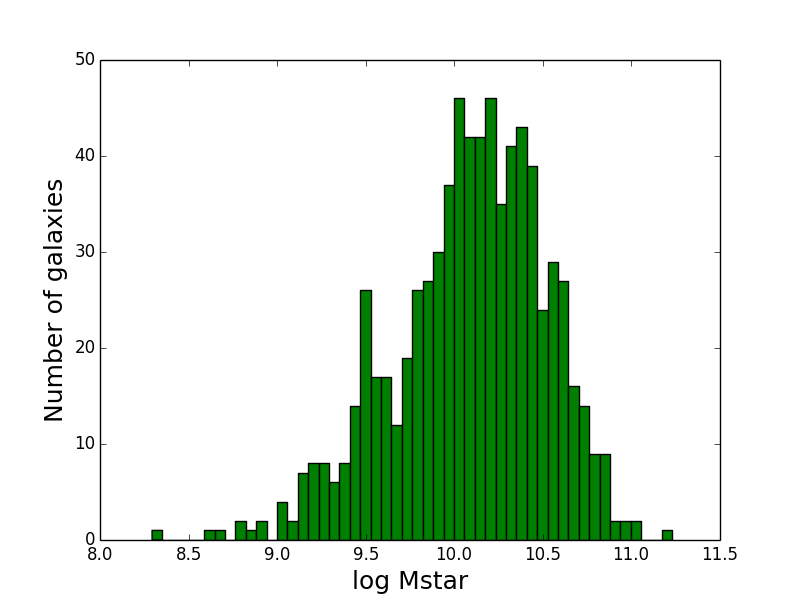}
\caption{Top: The redshift distribution of the UV-H$\alpha$-IR joint sample of 745 objects. The median redshift is $z=0.077$. Most objects in our sample are at $z<0.15$. Bottom: The stellar mass (in unit of $M_{\odot}$) distribution of the joint sample. The median stellar mass is $\log_{M_{\rm star}}=10.13M_{\odot}$.}
\label{fig:hist}
\end{figure}

\begin{figure}
\includegraphics[height=2.4in,width=3.4in]{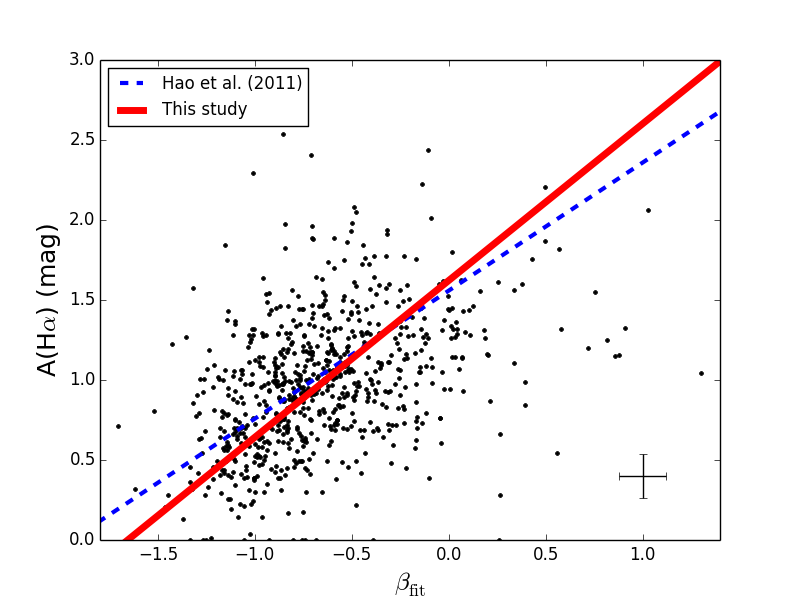}
\includegraphics[height=2.4in,width=3.4in]{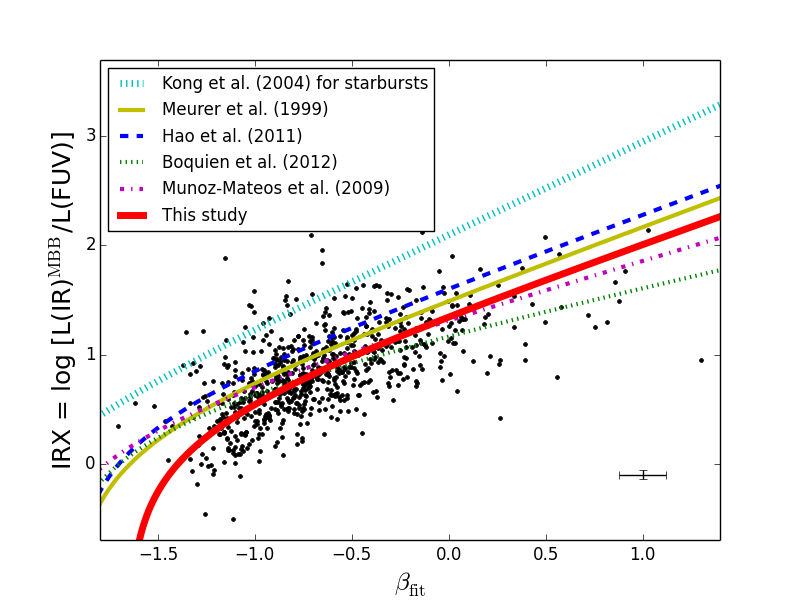}
\caption{Top: Attenuation in H$\alpha$ line derived from the Balmer decrement ratio A(H$\alpha$) vs UV spectral slope $\beta_{\rm fit}$ for galaxies in our joint UV-H$\alpha$-IR sample. The blue dashed line is the bisector fitting used in Hao et al. (2011). The red solid line is the bisector fitting to our sample. The black error bars indicate the median errors of the x- and y-axis.  Bottom: IRX (i.e. $\log_{10} L_{\rm IR}/L_{\rm FUV}$) vs $\beta_{\rm fit}$. The blue dashed line is from Hao et al. (2011). The green dotted line is derived for star-forming galaxies in the Herschel Reference Survey from Boquien et al. (2012). The magenta dash-dot line is derived for the SINGS sample from Mu{\~n}oz-Mateos et al. (2009). The thin yellow solid line is the relation for starburst galaxies from Meurer et al. (1999). The thick cyan dotted line is the best-fit relation for starbursts from Kong et al. (2004).  The thick red solid line is the best fit to our sample. The black error bars indicate the median errors of the x- and y-axis. The median statistical error on A(H$\alpha$), $\beta$ and IRX is 0.14, 0.12 and 0.04, respectively.}
\label{fig:dust_atten}
\end{figure}

\begin{figure}
\includegraphics[height=2.4in,width=3.4in]{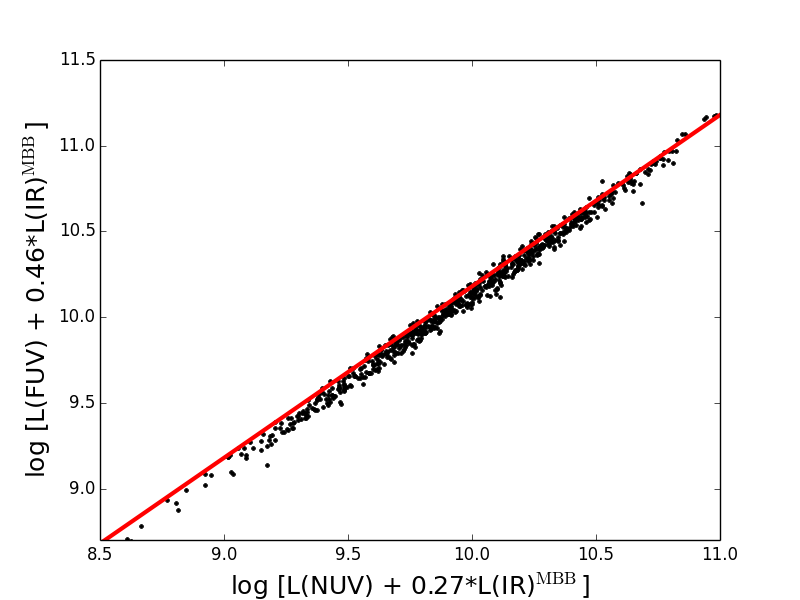}
\caption{$\log [{\rm L(FUV)} + 0.46*L({\rm IR})]$ vs $\log [{\rm L(NUV)} + 0.27*L({\rm IR})]$ for galaxies in our joint UV-H$\alpha$-IR sample. The correlation is very tight. The red line is the expected correlation based on matching the SFR prescriptions (i.e. Eq. 13 and Eq. 14).}
\label{fig:FUVIR_vs_NUVIR}
\end{figure}

\section{Correlation between different SFR indicators}
\subsection{Properties of the selected samples}

In the top panel in Fig.~\ref{fig:parent_vs_selected}, we compare the fraction of the selected samples at different wavelengths with the parent sample as a function of redshift. The construction of the parent sample is described in Section 2.1. The selection criteria of the UV, H$\alpha$ and IR samples are summarised in Table 3. The blue dots correspond to galaxies in the UV sample by requiring S/N $\geq10$ in both FUV and NUV and at least two filters in the rest-frame UV over $0.01<z<0.273$. The green dots correspond to galaxies in the selected H$\alpha$ sample by requiring S/N $\ge15$ in both H$\alpha$ and H$\beta$ emission lines for all star-forming galaxies over $0.01<z<0.273$ using the Kauffmann et al. (2003) classification. The red dots correspond to galaxies in the selected IR sample by requiring galaxies $0.01<z<0.273$ detected at S/N $\geq3$ at 22 and 250 $\mu$m and S/N $\geq3$ at either 100 or 160 $\mu$m. The UV selected sample contains more objects compared to the IR or the H$\alpha$ selection. Both the UV and the H$\alpha$ samples preferentially select lower-redshift objects, while the IR selected sample is better at picking up objects at higher redshifts compared to the UV or H$\alpha$ selection.  In the bottom panel in Fig.~\ref{fig:parent_vs_selected}, we show the fraction of the selected samples as a function of stellar mass. It is clear that the UV and H$\alpha$ selected samples preferentially select low-mass galaxies (related to the fact that the UV and H$\alpha$ selection favour galaxies at lower redshifts), while the completeness fraction of the IR-selected sample has a peak in the middle and declines toward both the low-mass and high-mass end. This peak can be explained by limited sensitivity in the IR (so the completeness at low mass, i.e. low SFR, is low) and increasingly higher passive fraction at high masses.

In the following sections, we will use a joint sample by applying the UV, H$\alpha$ and IR selection criteria together, which contains a total of 745 objects. Fig.~\ref{fig:hist} shows the full redshift distribution and stellar mass distribution of the UV-H$\alpha$-IR joint sample. We can see that the vast majority of our sources are at $z<0.15$. The median redshift of the joint sample is $z=0.077$ and the median stellar mass is $\log_{M_{\rm star}}=10.13M_{\odot}$.

\subsection{Dust attenuation}

Before we proceed to examine the correlation between different SFR indicators, we can look at the relation between different estimates of dust attenuation, i.e., the observed FUV - NUV colour or equivalently the UV spectral slope $\beta$, the total infrared to FUV luminosity ratio IRX (i.e. IRX$=\log[L_{\rm IR}/L_{\rm FUV}]$), and the attenuation in the H$\alpha$ line based on the measured Balmer decrement ratio.

In the top panel of Fig.~\ref{fig:dust_atten}, we show the attenuation in H$\alpha$ line based on the Balmer decrement ratio $A_{\rm H\alpha}$ vs the UV spectral slope $\beta$. The blue line is the bisector fitting to the Moustakas \& Kennicutt (2006) sample used in Hao et al. (2011),
\begin{equation}
A_{\rm H\alpha} = 0.8 \beta +1.6,
\end{equation}
or equivalently as a function of the observed FUV-NUV colour,
\begin{equation}
A_{\rm H\alpha} = 1.86 ({\rm FUV} - {\rm NUV})_{\rm obs} - 0.04.
\end{equation}
Hao et al. (2011) used the correlation between $\beta$ (or the observed FUV-NUV colour) and $A_{\rm H\alpha}$ to empirically determine the unattenuated power-law slope of the UV continuum or equivalently the intrinsic (dust-free) FUV-NUV colour, (FUV-NUV)$_{\rm int}$. From Eq. 16, when $A_{\rm H\alpha}=0$,  (FUV-NUV)$_{\rm int}=0.022$.
However, we find that the Hao et al. relation is not a good description of our sample. Instead, the red line is the bisector fitting to our sample in this paper, 
\begin{equation}
A_{\rm H\alpha} = 0.98\beta_{\rm fit} + 1.62.
\end{equation}
So, when setting $A_{\rm H\alpha}$ to 0, the dust-free (FUV-NUV)$_{\rm int}=0.15$ for our sample. If using $\beta_{\rm colour}$ instead of $\beta_{\rm fit}$, then 
\begin{equation}
A_{\rm H\alpha} = 0.91\beta_{\rm colour} + 1.49,
\end{equation}
and the dust-free (FUV-NUV)$_{\rm int}=0.16$.

In the bottom panel of Fig.~\ref{fig:dust_atten}, the total infrared to FUV luminosity ratio IRX (i.e. IRX$=\log[L_{\rm IR}/L_{\rm FUV}]$) is plotted against $\beta$. The blue line is from Hao et al. (2011), 
\begin{equation}
{\rm IRX} = \log \frac{[10^{0.4s_{\rm FUV}[({\rm FUV-NUV})_{\rm obs} - ({\rm FUV-NUV})_{\rm int}]} - 1] }{a_{\rm FUV}},
\end{equation}
where the slope of the UV part of the attenuation curve $s_{\rm FUV}=3.83$ and the fraction of the total infrared luminosity due to recent star formation $a_{\rm FUV}=0.46$. For comparison, the green line is from Boquien et al. (2012) which studied the IRX-$\beta$ relation on sub galactic scales in star-forming galaxies selected from the Herschel Reference Survey. The magenta line is the best-fit IRX-$\beta$ relation for normal star-forming spiral galaxies in the SINGS sample from Mu{\~n}oz-Mateos et al. (2009). The yellow line is well-known Meurer et al. (1999) relation for starburst galaxies. The cyan line is best-fit IRX-$\beta$ relation for nearby starbursts in Kong et al. (2004). The red line is the best-fit (of the functional form defined in Eq. 19) to our sample with $s_{\rm FUV}=3.67$ and $a_{\rm FUV}=0.46$. If using $\beta_{\rm colour}$ instead of $\beta_{\rm fit}$, then we find $s_{\rm FUV}=3.55$ and $a_{\rm FUV}=0.46$ for our sample. Note that in our fitting procedure, we have fixed $a_{\rm FUV}$ at 0.46. This is because the correlation between IRX and $\beta$ is relatively poor so it is advantageous to minimise the number of free parameters. In addition, we believe $a_{\rm FUV}=0.46$ is suitable value for our sample, i.e. it corresponds to approximately the correct fraction of the total infrared luminosity that is produced by recent star formation. For example, Fig.~ \ref{fig:FUVIR_vs_NUVIR} shows $\log [{\rm L(FUV)} + 0.46*L({\rm IR})]$ is extremely well and tightly correlated with $\log [{\rm L(NUV)} + 0.27*L({\rm IR})]$ for galaxies in our joint sample. Further evidence comes from the good correlation between the UV + IR based SFR indicator and the H$\alpha$ based SFR indicator (see Fig.~\ref{fig:SFR_comp_ha} in Section 4.3). To further understand the differences in the IRX-$\beta$ relation between the Hao et al. (2011) study and this paper, we investigate in detail the galaxy samples used in both studies in Appendix~\ref{appendix2}. We show that the difference between the the Hao et al. (2011) relation based on a nearby star-forming sample and the new relation derived in this paper is due to the difference in the galaxy samples. In addition to being at higher redshifts, our galaxy sample corresponds to much lower survey flux limits (by more than an order of magnitude) in the IR and UV and therefore contains many more quiescent star-forming galaxies with redder UV spectral slopes and lower IRX values.

\begin{figure}
\includegraphics[height=2.4in,width=3.4in]{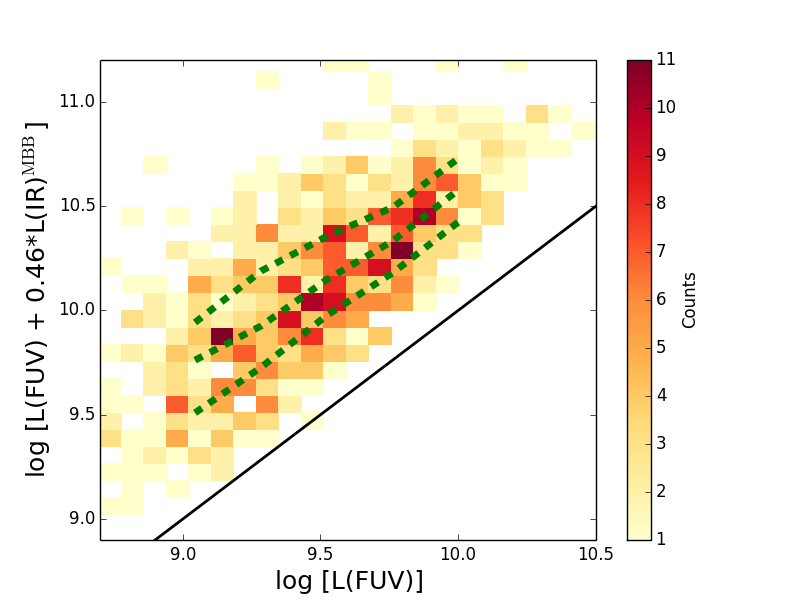}
\includegraphics[height=2.4in,width=3.4in]{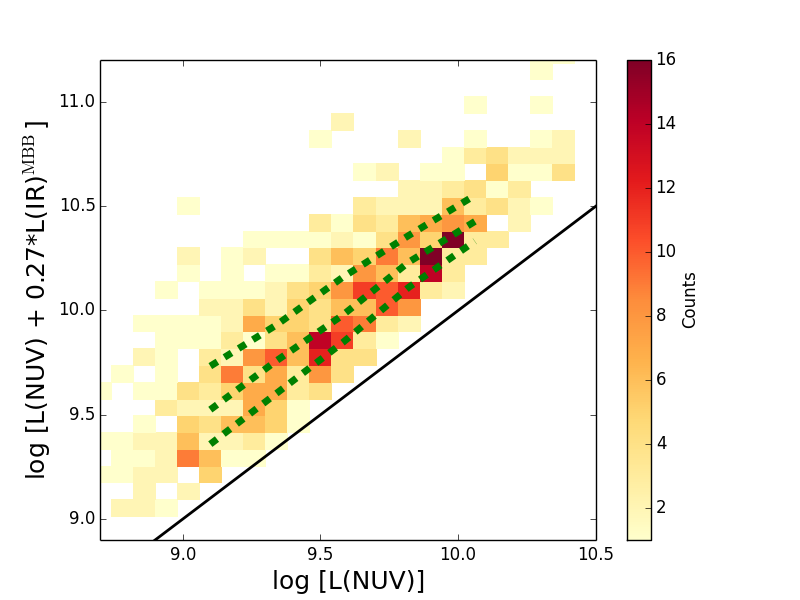}
\caption{Top:  The linear combination of the observed FUV luminosity and total infrared luminosity vs the observed FUV luminosity (uncorrected for dust attenuation), colour-coded by galaxy counts. The black solid line is the predicted relation from matching the SFR prescriptions (i.e. Eq. 6 and Eq. 13). The green dashed lines mark the 25th, 50th and 75th percentile.  Bottom:  The linear combination of the observed NUV luminosity and total infrared luminosity vs the observed NUV luminosity (uncorrected for dust attenuation), colour-coded by galaxy counts. The black solid line is the predicted relation from matching the SFR prescriptions (i.e. Eq. 7 and Eq. 14). The green dashed lines mark the 25th, 50th and 75th percentile. }
\label{fig:LUV_uncorr_vs_UVIR}
\end{figure}

\begin{figure}
\includegraphics[height=2.4in,width=3.4in]{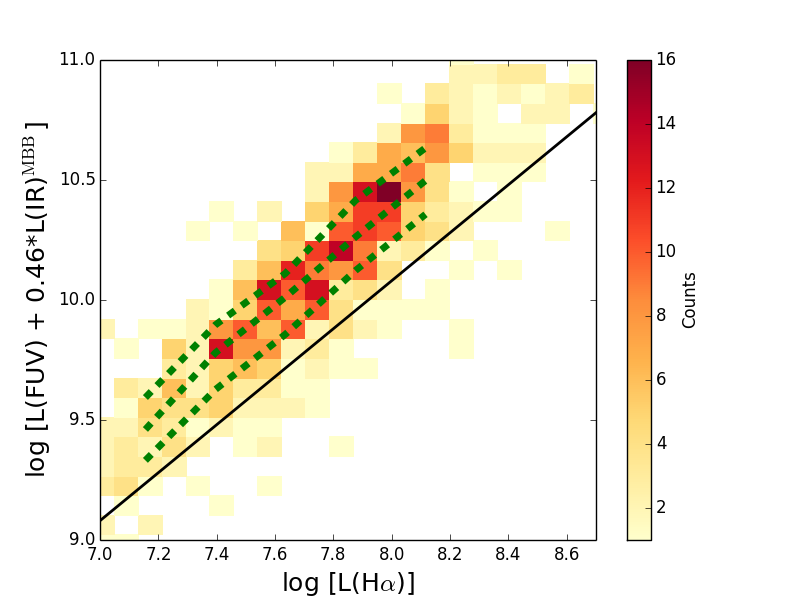}
\includegraphics[height=2.4in,width=3.4in]{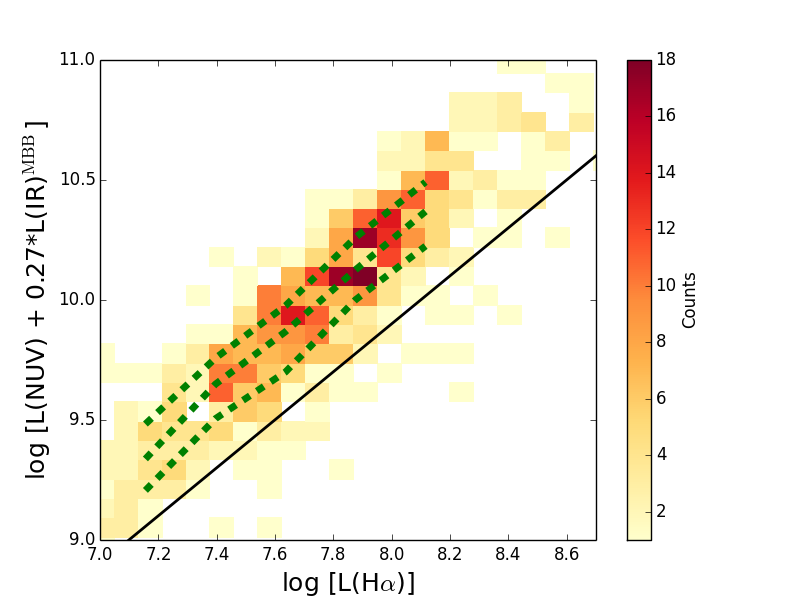}
\caption{Top: The linear combination of the observed FUV luminosity and total infrared luminosity vs the observed H$\alpha$ line luminosity (uncorrected for dust attenuation), colour-coded by galaxy counts. The black line is the predicted relation from matching the SFR prescriptions (i.e. Eq. 9 and Eq. 13). The green dashed lines mark the 25th, 50th and 75th percentile.  Bottom:  The linear combination of the observed NUV luminosity and total infrared luminosity vs the observed H$\alpha$ line luminosity (uncorrected for dust attenuation), colour-coded by galaxy counts. The black line is the predicted relation from matching the SFR prescriptions (i.e. Eq. 9 and Eq. 14). The green dashed lines mark the 25th, 50th and 75th percentile. }
\label{fig:LHa_uncorr_vs_UVIR}
\end{figure}

\begin{figure}
\includegraphics[height=2.4in,width=3.4in]{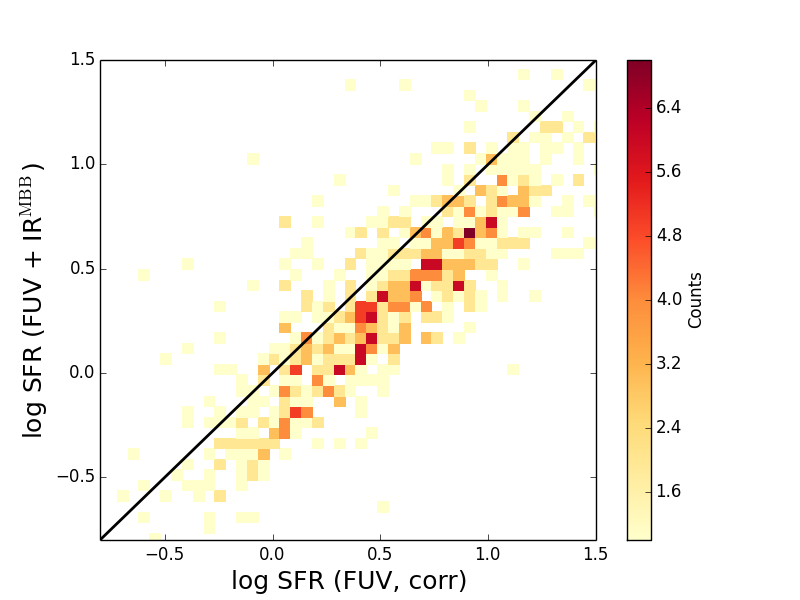}
\includegraphics[height=2.4in,width=3.4in]{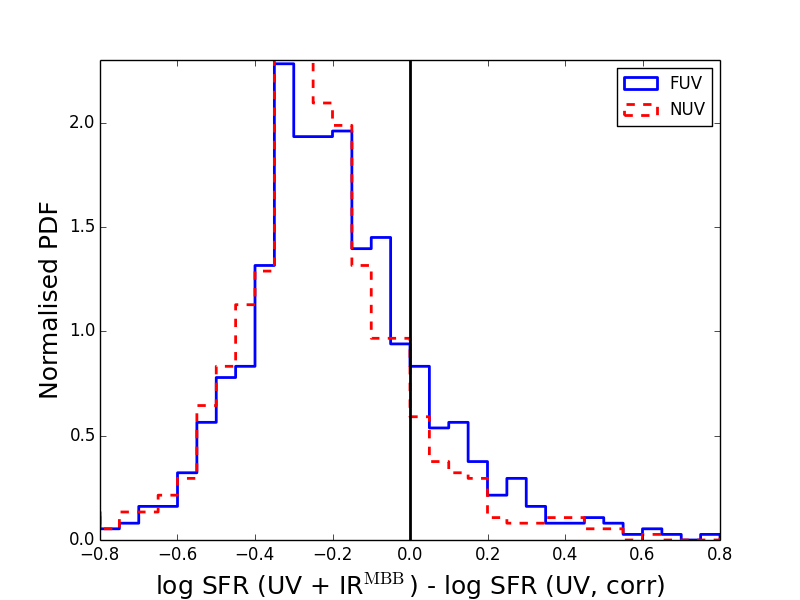}
\caption{Top: ${\rm SFR}_{\rm FUV + IR}$ (derived using Eq. 13) vs ${\rm SFR}_{\rm FUV, corr}$ (derived using Eq. 6), in unit of $M_{\odot}$/yr, colour-coded by galaxy counts. The ${\rm SFR}_{\rm FUV, corr}$  values are derived using $\beta_{\rm fit}$ and the Hao et al. (2011) relation between $A_{\rm FUV}$ and $\beta$ (i.e. Eq. 4). The black solid line is the one-to-one relation. Bottom: The normalised histogram in the difference between the UV+IR SFR indicator and the UV SFR indicator corrected for attenuation using $\beta$ (blue solid : ${\rm SFR}_{\rm FUV + IR} - {\rm SFR}_{\rm FUV, corr}$; red dashed : ${\rm SFR}_{\rm NUV + IR} - {\rm SFR}_{\rm NUV, corr}$). The black vertical line corresponds to ${\rm SFR}_{\rm UV + IR}={\rm SFR}_{\rm UV, corr}$.}
\label{fig:SFR_comp_uv}
\end{figure}

\begin{figure}
\includegraphics[height=2.4in,width=3.4in]{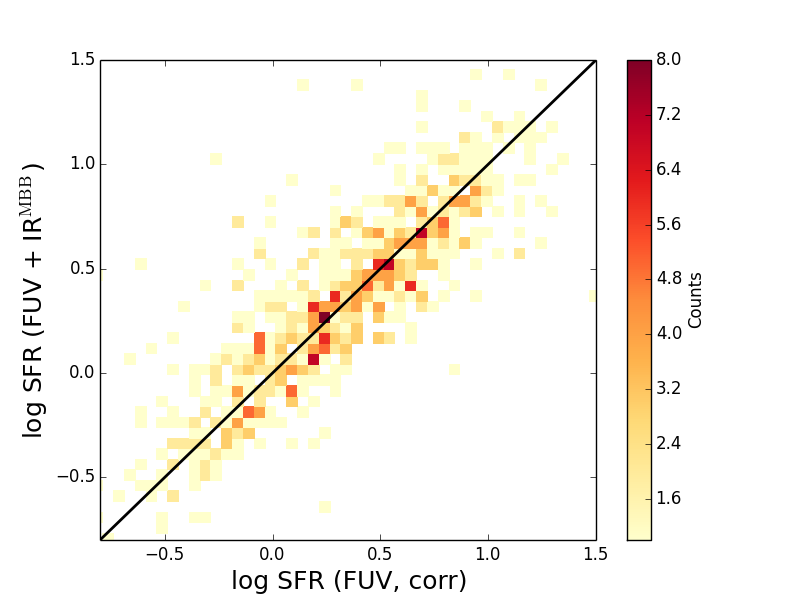}
\includegraphics[height=2.4in,width=3.4in]{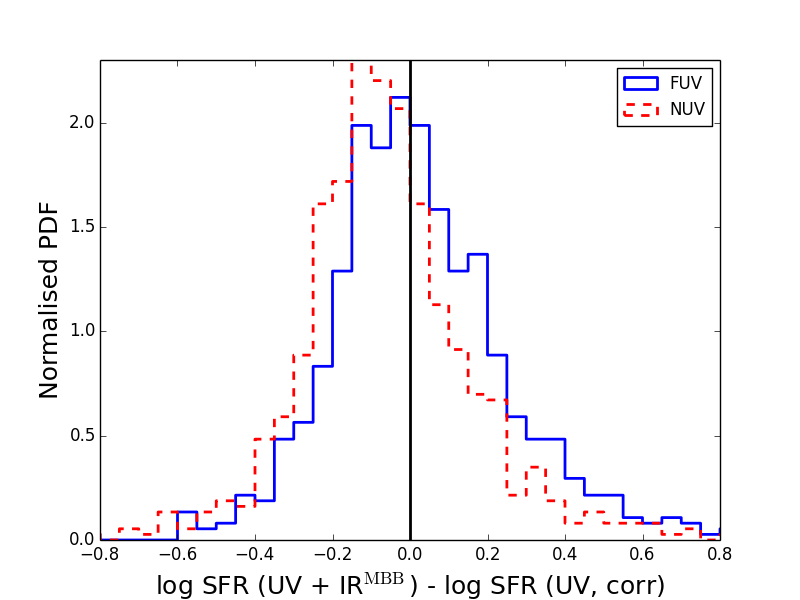}
\caption{Similar to Fig.~\ref{fig:SFR_comp_uv}, but the ${\rm SFR}_{\rm FUV, corr}$  values are derived using the new relation between $A_{\rm FUV}$ and $\beta$ (i.e. Eq. 21) for our joint  UV-H$\alpha$-IR sample.}
\label{fig:new_SFR_comp_uv}
\end{figure}

\begin{figure}
\includegraphics[height=2.4in,width=3.4in]{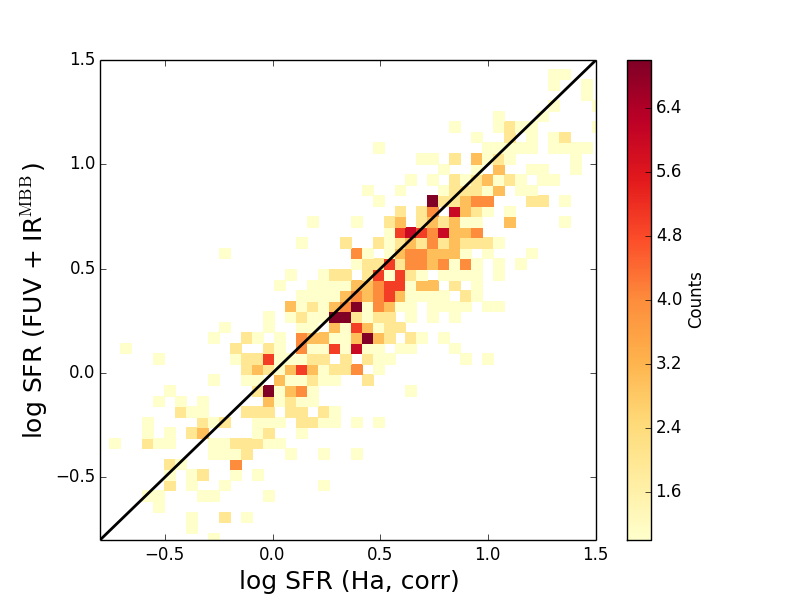}
\includegraphics[height=2.4in,width=3.4in]{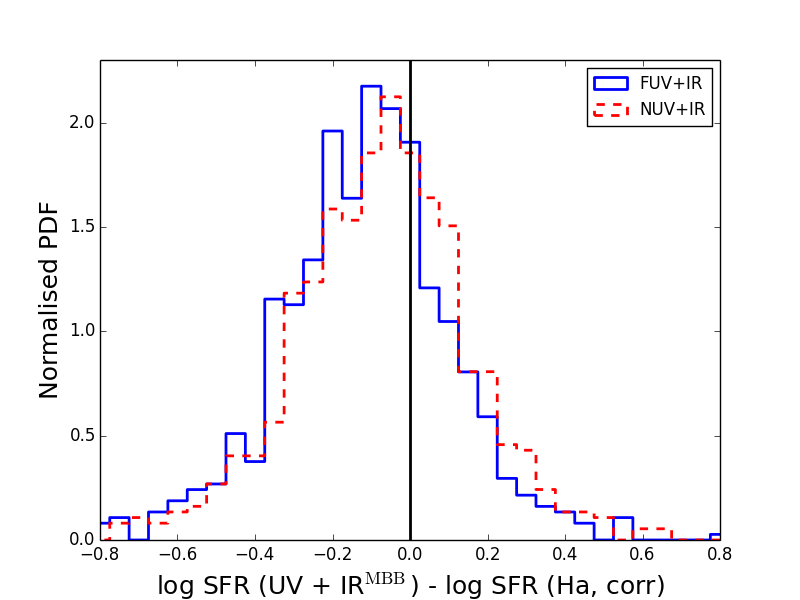}
\caption{Top: Comparison of ${\rm SFR}_{\rm H\alpha, corr}$ with ${\rm SFR}_{\rm FUV + IR}$ (derived using Eq. 13), in unit of $M_{\odot}$/yr, colour-coded by galaxy counts. The black solid line is the one-to-one relation. Bottom: The normalised histogram in the difference between the UV+IR SFR indicator and the H$\alpha$ SFR indicator corrected for attenuation using Balmer decrement  (blue solid: ${\rm SFR}_{\rm FUV + IR} - {\rm SFR}_{\rm H\alpha, corr}$; red dashed: ${\rm SFR}_{\rm NUV + IR} - {\rm SFR}_{\rm H\alpha, corr}$). The black vertical line corresponds to ${\rm SFR}_{\rm H\alpha, corr}={\rm SFR}_{\rm UV + IR}$.}
\label{fig:SFR_comp_ha}
\end{figure}

Following Hao et al. (2011), the FUV attenuation can be estimated from the observed (FUV-NUV) colour using 
\begin{equation}
A_{\rm FUV} = s_{\rm FUV} [({\rm FUV-NUV})_{\rm obs} - ({\rm FUV-NUV})_{\rm int})].
\end{equation}
So substituting the values for $s_{\rm FUV}$ and $({\rm FUV-NUV})_{\rm int})$ for our joint UV-H$\alpha$-IR sample into Eq. 19, we have
\begin{equation}
A_{\rm FUV} = 3.67 [({\rm FUV-NUV})_{\rm obs} - 0.15],
\end{equation}
or equivalently as a function of the UV spectral slope $\beta$, 
\begin{equation}
A_{\rm FUV} = 1.58\times \beta_{\rm fit} + 2.62.
\end{equation}
If using $\beta_{\rm colour}$ instead of $\beta_{\rm fit}$, then 
\begin{equation}
A_{\rm FUV} = 1.53\times \beta_{\rm colour} + 2.50.
\end{equation}

\subsection{SFR correlations}

First of all, we examine how the observed UV luminosities (FUV and NUV) and the H$\alpha$ line luminosity, both uncorrected for dust attenuation,  compare with the linear combination of the observed UV luminosity and the total infrared luminosity. Fig.~\ref{fig:LUV_uncorr_vs_UVIR} compares the linear combination of the observed UV luminosity and the total infrared luminosity with the observed UV luminosity (without correction for dust attenuation). The solid line in each panel is the predicted relation based on matching SFR prescriptions.  As expected, the FUV band suffers more from dust attenuation than the NUV. However, in the luminosity range probed by our sample, we do not see a significant difference in the amount of attenuation with changing UV luminosity. Fig.~\ref{fig:LHa_uncorr_vs_UVIR} compares the linear combination of the observed UV luminosity and the total infrared luminosity with the observed H$\alpha$ line luminosity (without correction for dust attenuation). Again, the solid line in each panel is the predicted relation based on matching SFR prescriptions. The effects of dust obscuration is less severe in the optical emission lines than in the UV. We see a similar amount of attenuation in the observed H$\alpha$ line luminosity in the top and bottom panel, as expected. Unlike Fig.~\ref{fig:LUV_uncorr_vs_UVIR}, here we do see some evidence that more actively star-forming galaxies suffer more from dust attenuation. 

\begin{table*}
\caption{The 16th, 50th and 84th percentile in the difference  of SFR indicators $\Delta {\rm SFR}={\rm SFR_{UV+IR^{\rm MBB}}} - {\rm SFR_{UV, corr}}$, where UV means either FUV or NUV. The dust attenuation correction applied to the UV SFR indicator is based on the UV spectral slope $\beta$. We compare the difference in SFR indicators using $\beta_{\rm fit}$ or $\beta_{\rm colour}$. We also compare the difference between applying the Hao et al. (2011) $A_{\rm FUV}$-$\beta$ relation and applying the new $A_{\rm FUV}$-$\beta$  relations (Eq. 22 and Eq. 23) derived in this paper.}\label{table:beta}
\begin{tabular}{lll}
\hline
&Hao et al. relation  & New relation in this paper   \\
\hline
SFR$_{\rm FUV+IR^{MBB}}$ - SFR$_{\rm FUV, corr, \beta_{\rm fit}}$        &  -0.4, -0.2, 0.0 &    -0.2, 0.0, 0.2   \\
\hline
SFR$_{\rm FUV+IR^{MBB}}$ - SFR$_{\rm FUV, corr, \beta_{\rm colour}}$ &  -0.5, -0.3, 0.0  &-0.2, 0.0, 0.2\\
\hline 
SFR$_{\rm NUV+IR^{MBB}}$ - SFR$_{\rm NUV, corr, \beta_{\rm fit}}$ &  -0.4, -0.3, -0.1  &-0.2, -0.1, 0.1\\
\hline
SFR$_{\rm NUV+IR^{MBB}}$ - SFR$_{\rm NUV, corr, \beta_{\rm colour}}$ & -0.5, -0.3, -0.1 &-0.3, -0.1, 0.1\\
\hline
\end{tabular}
\end{table*}

Now we can examine how different SFR indicators compare after applying corrections for dust attenuation. Starting with the UV SFR indicators, we apply dust correction factors based on the UV spectral slope  $\beta$ (or equivalently the observed (FUV-NUV) colour) using Eq. 4 in Section 3.1. In Fig.~\ref{fig:SFR_comp_uv}, we show comparisons between SFR$_{\rm UV, corr}$ (based on dust corrected UV luminosity using $\beta$) and SFR$_{\rm UV + IR}$. There is a good correlation (linear and tight) between SFR$_{\rm UV, corr}$ and SFR$_{\rm UV + IR}$, but there is a large overall shift (around 0.3 dex). SFR$_{\rm UV, corr}$ gives systematically higher values than SFR$_{\rm UV + IR}$. The offset is likely to be caused by an incorrect $A_{\rm FUV}$-$\beta$ relation. We have demonstrated in Section 4.2 that the Hao et al. (2011) $A_{\rm FUV}$-$\beta$ relation does not provide a suitable description for our sample. So, we apply the new $A_{\rm FUV}$-$\beta$ relation (i.e. Eq. 22) to derive SFR$_{\rm UV, corr}$ which are plotted in Fig.~\ref{fig:new_SFR_comp_uv}.  It is clear that large offset  between SFR$_{\rm UV, corr}$ and SFR$_{\rm UV + IR}$ has now gone away.  In Table 4, we list the 16th, 50th and 84th percentile in the difference between SFR$_{\rm UV, corr}$ and SFR$_{\rm UV + IR}$, where UV means either FUV or NUV.  We compare the difference in $\Delta {\rm SFR}$ using $\beta_{\rm fit}$ or $\beta_{\rm colour}$. We also compare the difference between applying the Hao et al. (2011) $A_{\rm FUV}$-$\beta$ relation and applying the new $A_{\rm FUV}$-$\beta$  relations (Eq. 22 and Eq. 23) derived in this paper. The percentiles are similar regardless of using $\beta_{\rm fit}$ or $\beta_{\rm colour}$ and regardless of using the FUV or NUV band.  There is still a small offset (around 0.1 dex) between SFR$_{\rm NUV, corr}$ and SFR$_{\rm NUV + IR}$, using either $\beta_{\rm fit}$ or $\beta_{\rm colour}$. This is likely due to the assumption made on the relation between $A_{\rm FUV}$ and $A_{\rm NUV}$ (see Section 3.1).

In Fig.~\ref{fig:SFR_comp_ha}, we show comparisons between SFR$_{\rm H\alpha, corr}$ and SFR$_{\rm UV + IR}$.  The dust attenuation corrections applied in SFR$_{\rm H\alpha, corr}$ are based on the Balmer decrement measurements. Again, the correlations are similar whether using FUV or NUV. SFR$_{\rm H\alpha, corr}$ gives systematically higher values than SFR$_{\rm UV + IR}$.  In Table 5, we list the 16th, 50th and 84th percentile in the difference  of SFR indicators $\Delta {\rm SFR}={\rm SFR_{UV+IR}} - {\rm SFR_{H\alpha, corr}}$. We compare the difference between using the infrared luminosity $L_{\rm IR}$ derived from the modified blackbody (MBB) template library and using $L_{\rm IR}$ derived from the Dale \& Helou (2002) (DH) template library. It is clear that the small systematic offset between SFR$_{\rm H\alpha, corr}$ and SFR$_{\rm UV + IR}$ see in Fig. 13 can be entirely explained by systematic error in $L_{\rm IR}$\footnote{Rosario et al. (2016) report excellent agreement between H$\alpha$-based SFR and SFR$_{\rm UV + IR}$ using their SDSS-Herschel matched sample. They use the Dale \& Helou (2002) library in deriving $L_{\rm IR}$.}.  The offset could also be due to other systematic errors of a similar scale in the $H\alpha$-based SFR tracer, e.g., optical depth effects in the Balmer corrections, systematic error in Balmer line absorption corrections or aperture effects, etc.

In Appendix~\ref{appendix3}, we further compare the three empirical SFR indicators considered in this paper, i.e., SFR$_{\rm UV, corr}$, SFR$_{\rm H\alpha, corr}$ and SFR$_{\rm UV + IR}$, with SFRs from Grootes et al. (2013) which are derived using the radiative transfer models of Popescu et al. (2011).

\begin{table}
\caption{The 16th, 50th and 84th percentile in the difference  of SFR indicators $\Delta {\rm SFR}={\rm SFR_{UV+IR}} - {\rm SFR_{H\alpha, corr}}$, where UV means either FUV or NUV. The dust attenuation correction applied to the H$\alpha$ SFR indicator is based on the Balmer decrement. We compare the difference between using the infrared luminosity $L_{\rm IR}$ derived from the modified blackbody (MBB) template library and using $L_{\rm IR}$ derived from the Dale \& Helou (2002) (DH) template library.}\label{table:beta}
\begin{tabular}{ll}
\hline
SFR$_{\rm FUV+IR^{\rm MBB}}$ - SFR$_{\rm H\alpha, corr}$ &  -0.3, -0.1, 0.1 \\
\hline
SFR$_{\rm FUV+IR^{\rm DH}}$ - SFR$_{\rm H\alpha, corr}$& -0.3, -0.0, 0.1 \\
\hline 
SFR$_{\rm NUV+IR^{\rm MBB}}$ - SFR$_{\rm H\alpha, corr}$  &-0.3, -0.1, 0.1\\
\hline
SFR$_{\rm NUV+IR^{\rm DH}}$ - SFR$_{\rm H\alpha, corr}$  &-0.2, -0.0, 0.2\\
\hline 
\end{tabular}
\end{table}

\subsection{Dependence on galaxy physical parameters}

\begin{figure*}
\includegraphics[height=2.2in,width=3.4in]{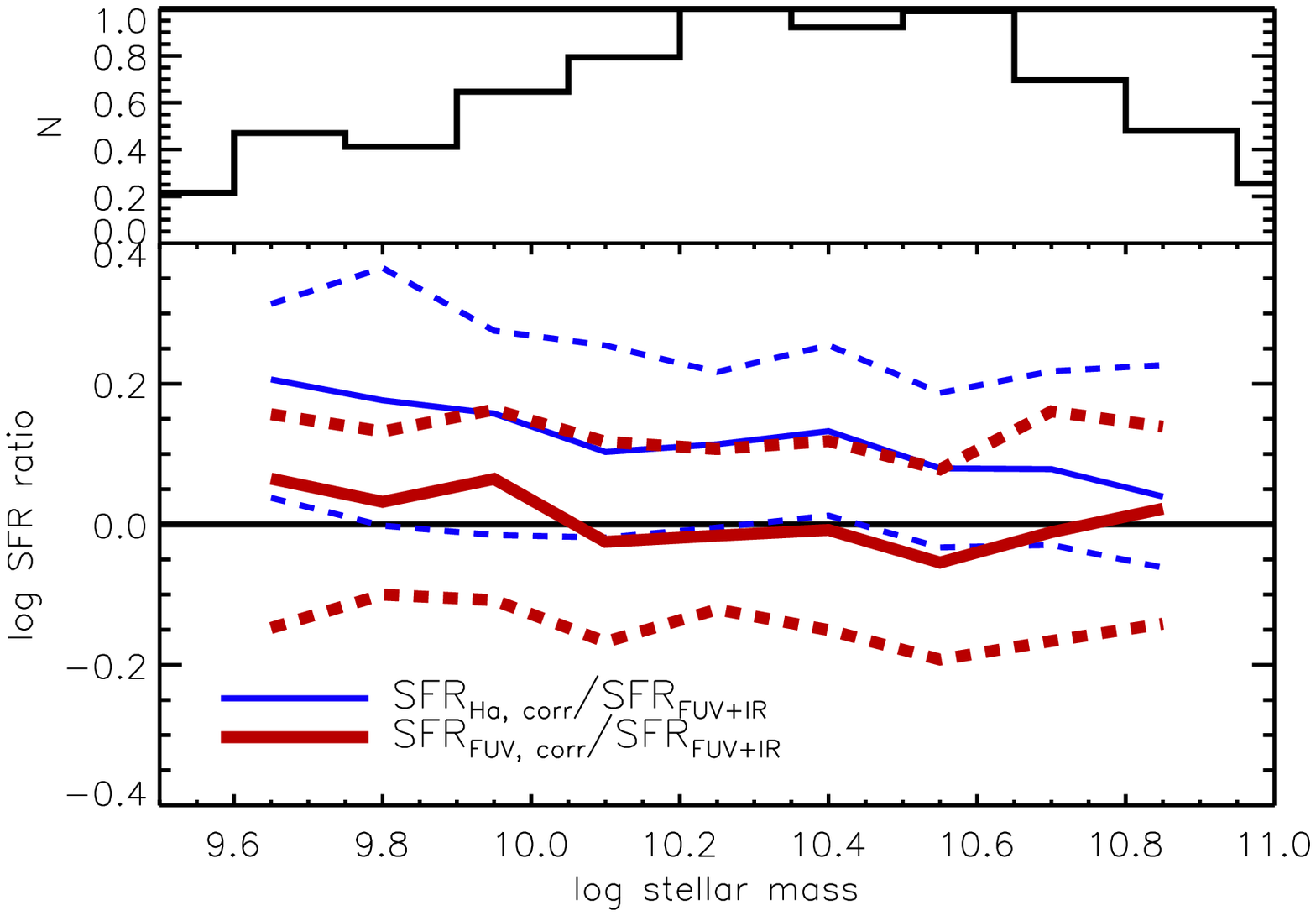}
\includegraphics[height=2.2in,width=3.4in]{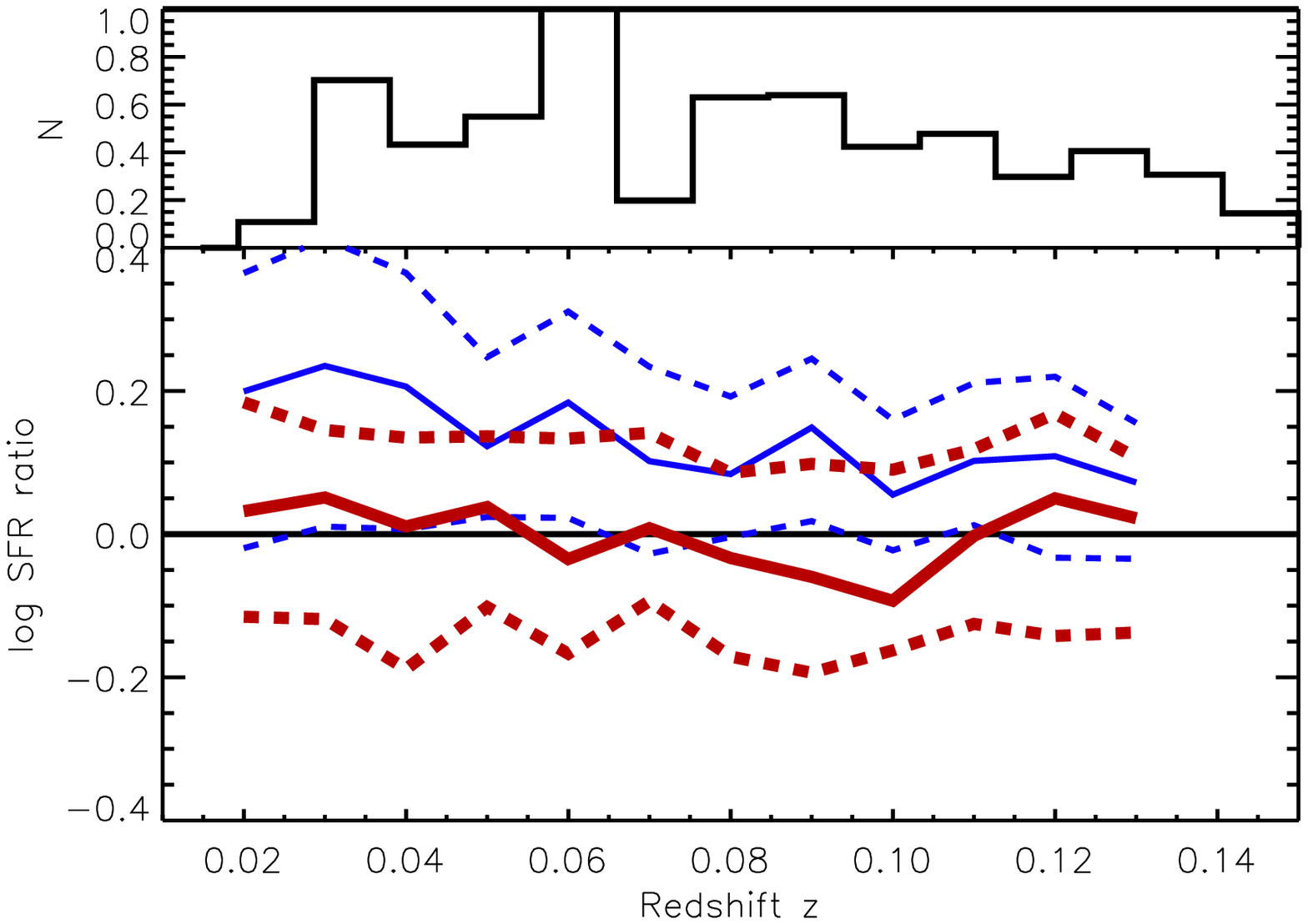}
\includegraphics[height=2.2in,width=3.4in]{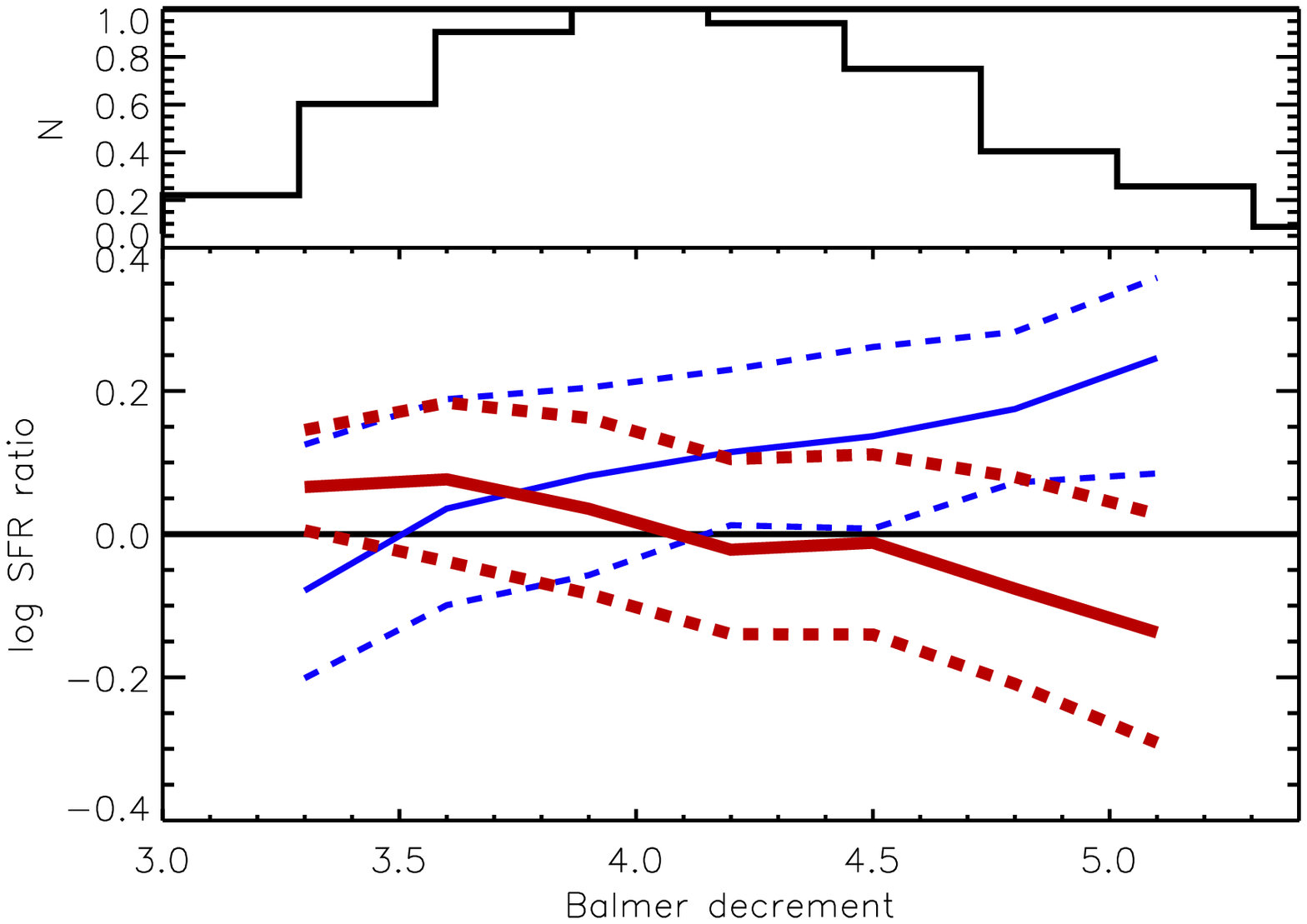}
\includegraphics[height=2.2in,width=3.4in]{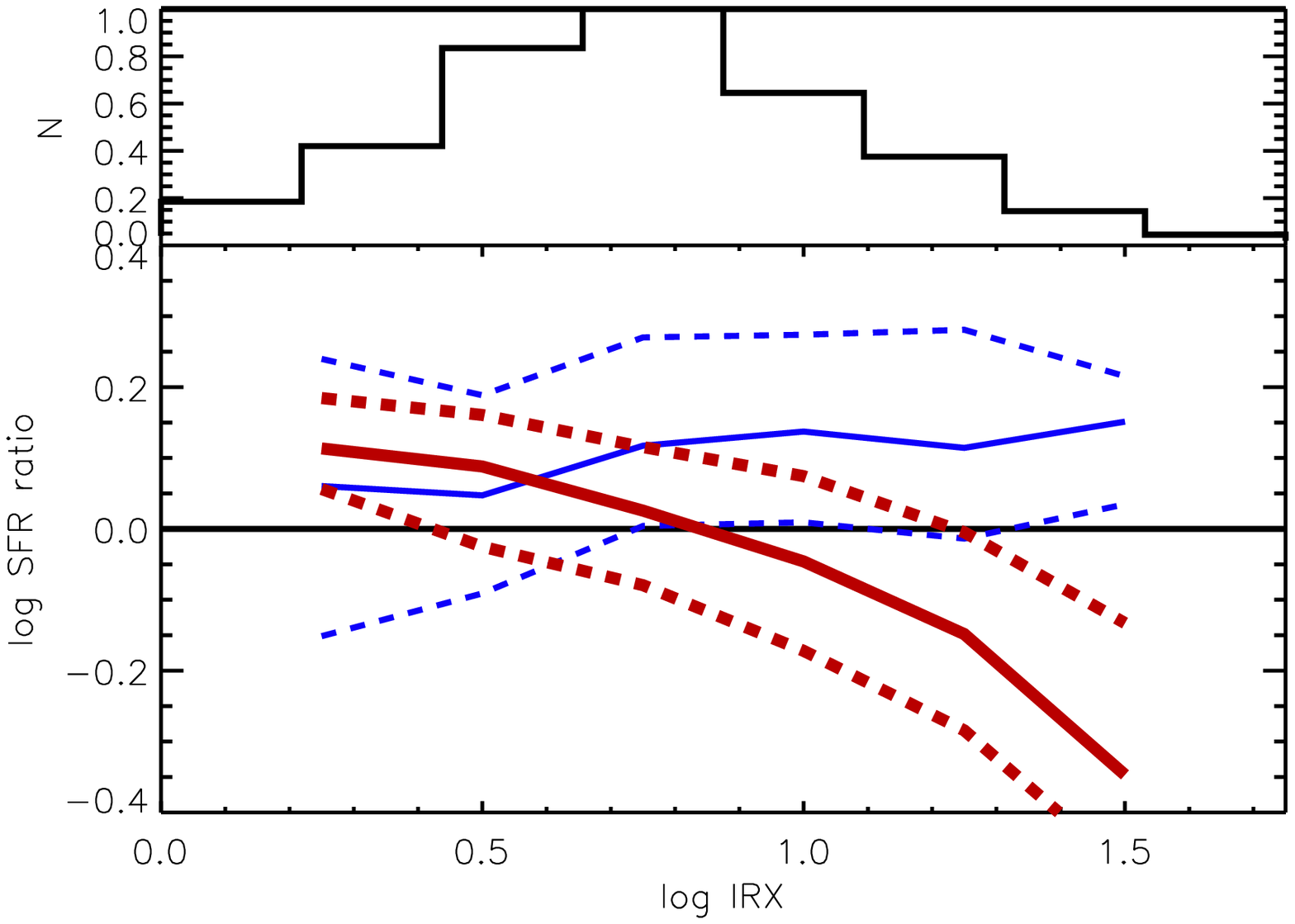}
\includegraphics[height=2.2in,width=3.4in]{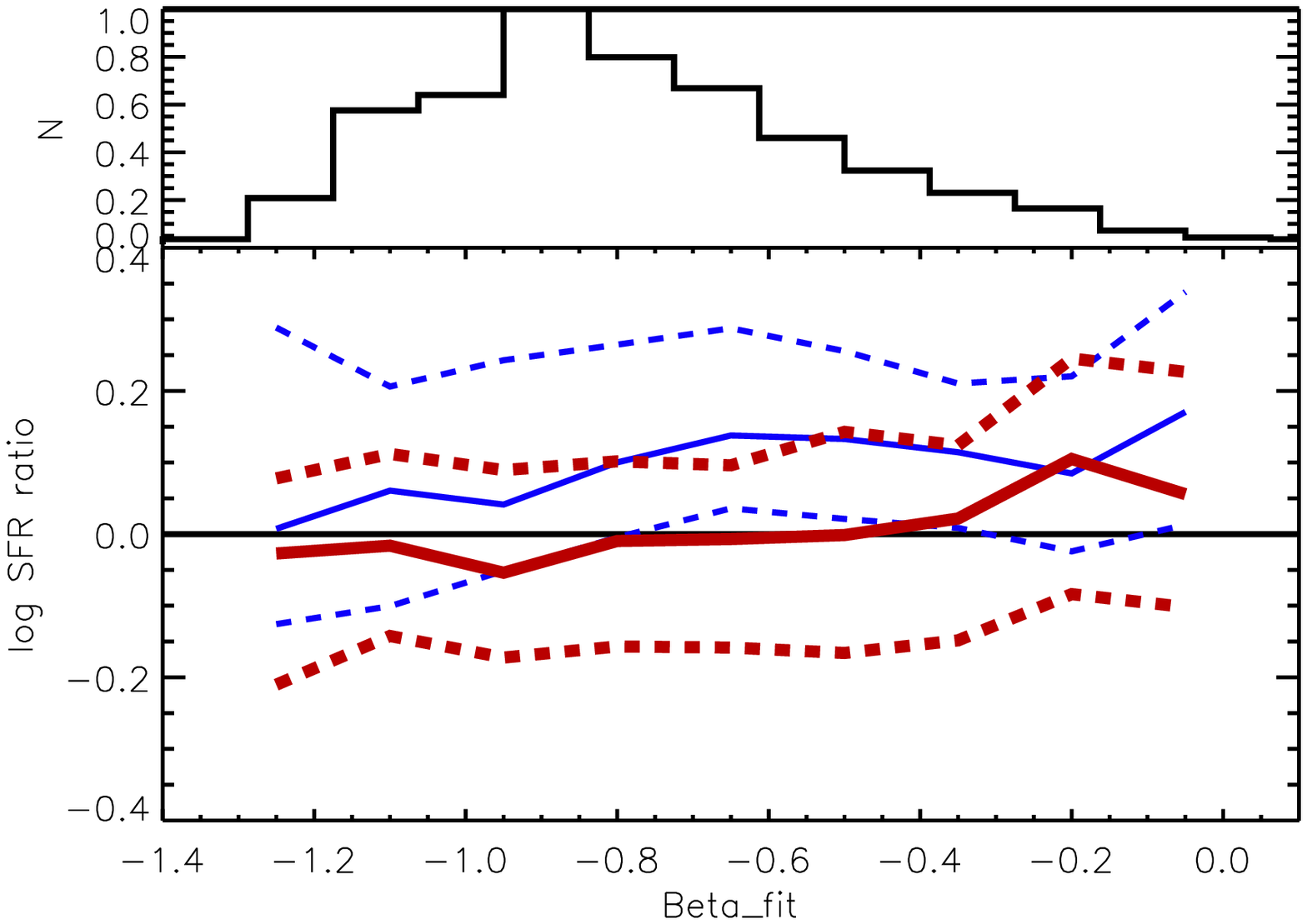}
\includegraphics[height=2.2in,width=3.34in]{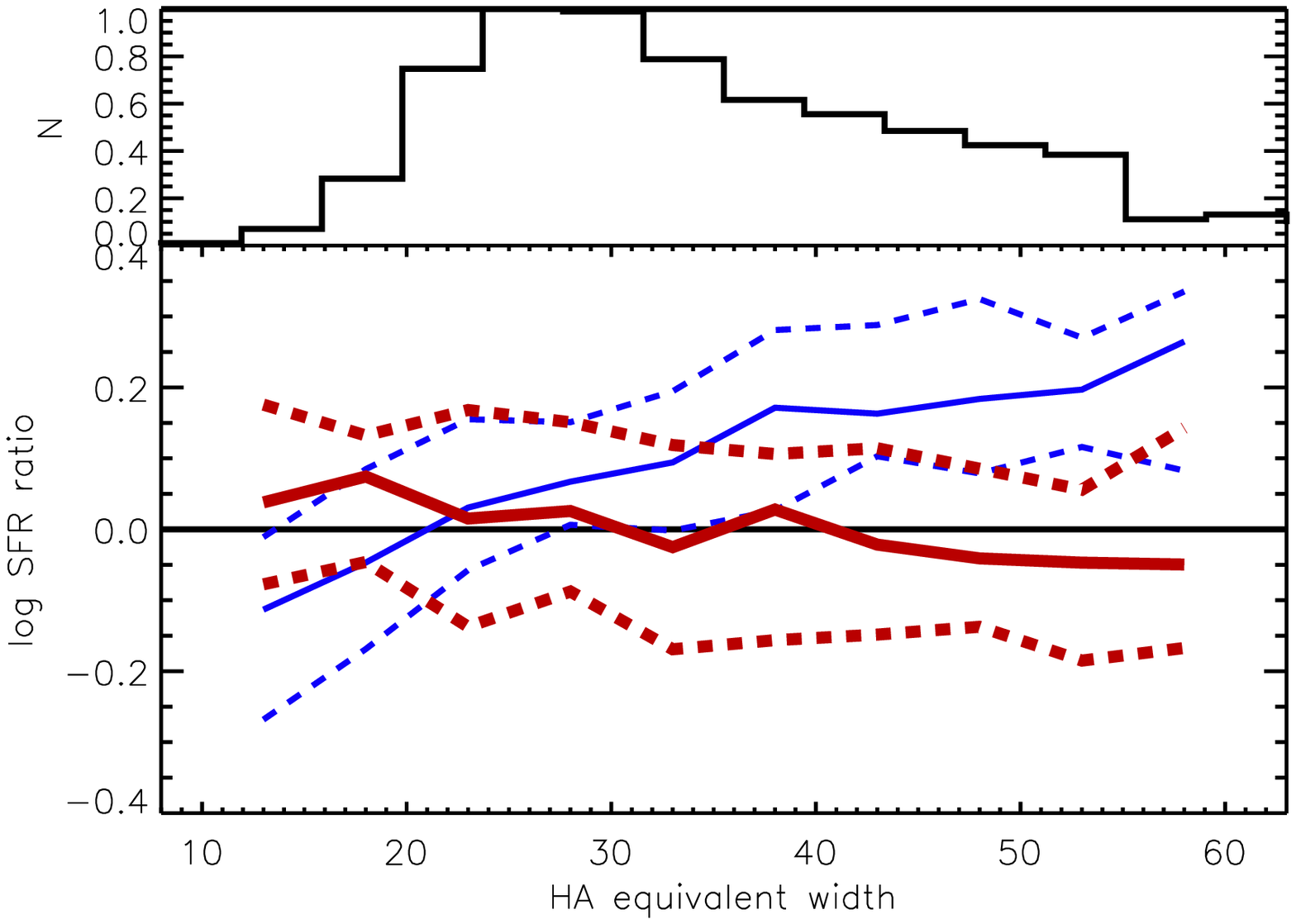}
\includegraphics[height=2.2in,width=3.34in]{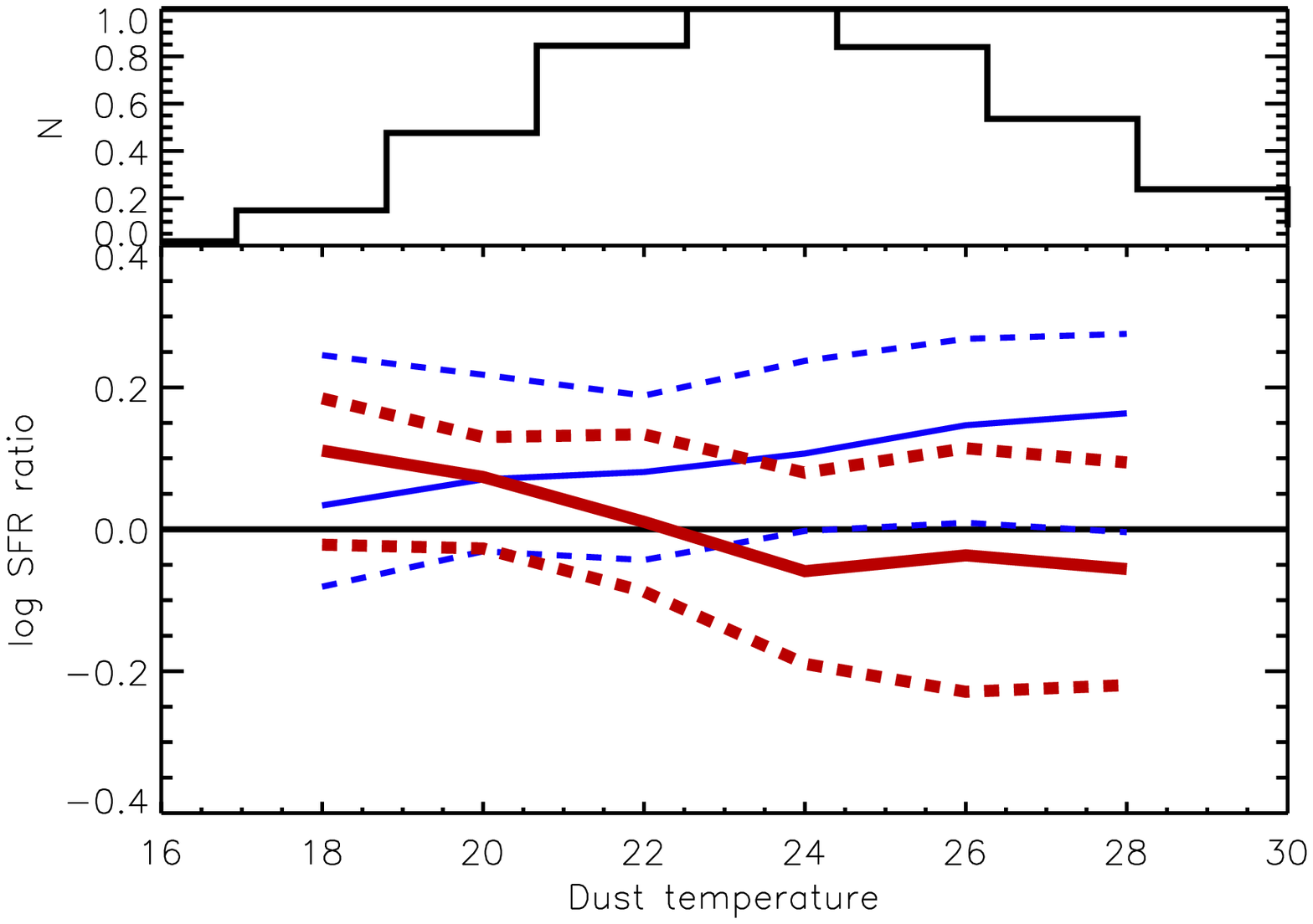}
\includegraphics[height=2.2in,width=3.34in]{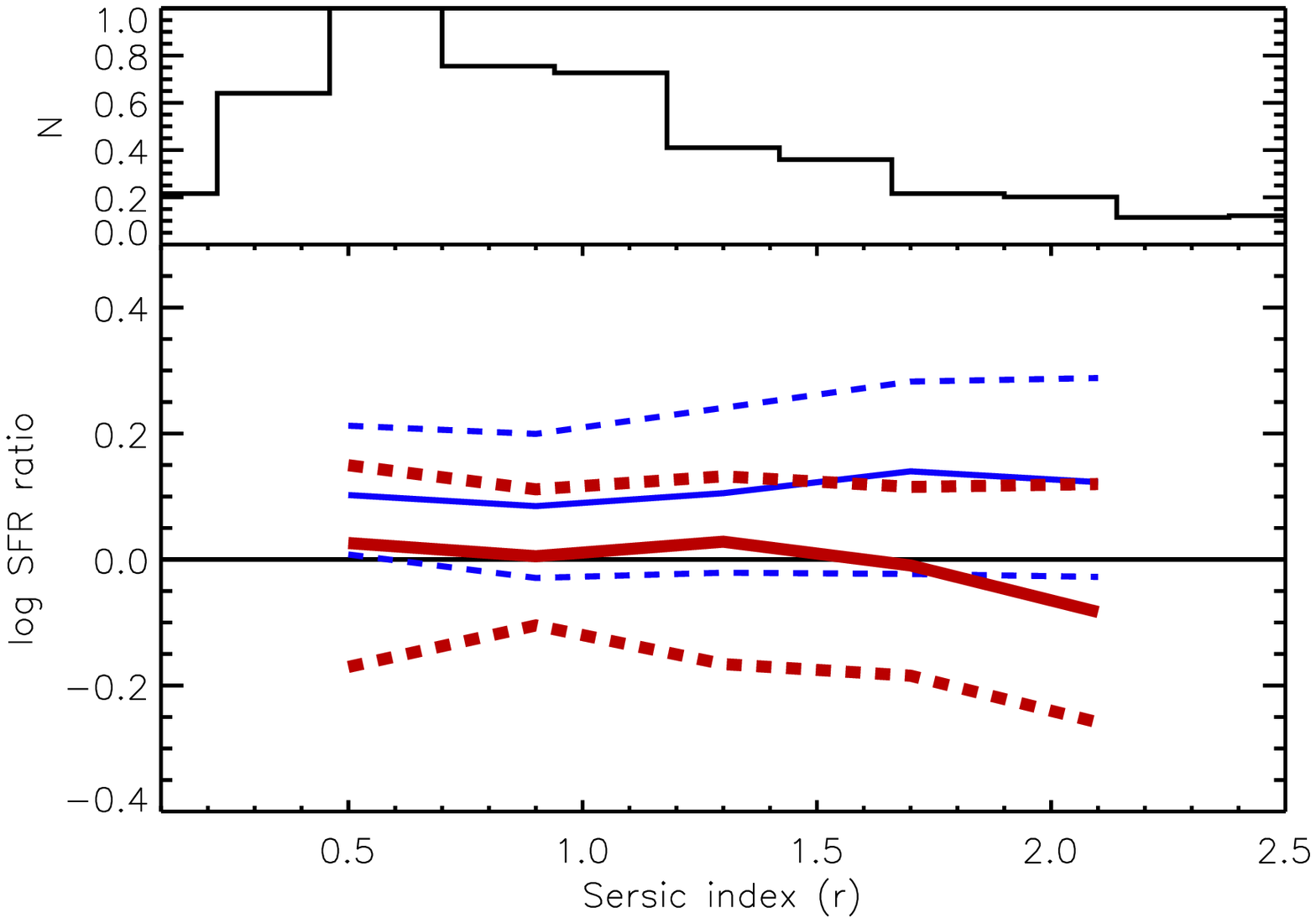}
\caption{The ratio of SFR indicators as a function of various physical parameters (stellar mass, redshift, Balmer decrement, IRX, $\beta_{\rm fit}$, H$\alpha$ equivalent width,  dust temperature, and Sersic index in the r-band). In the top panel of each figure, we show the normalised histogram of the x-axis.  The thin blue lines correspond to the 25th, 50th and 75th percentile in the ratio of SFR$_{\rm H\alpha, corr}$ over SFR$_{\rm FUV + IR}$. The thick red lines correspond to the 25th, 50th and 75th percentile in the ratio of SFR$_{\rm FUV, corr}$ (using $\beta_{\rm fit}$ and our new $A_{\rm FUV}$-$\beta$ relation Eq. 22) over SFR$_{\rm FUV + IR}$.}
\label{fig:sfr_comp_3rdvar}
\end{figure*}

\begin{figure*}
\includegraphics[height=2.2in,width=3.4in]{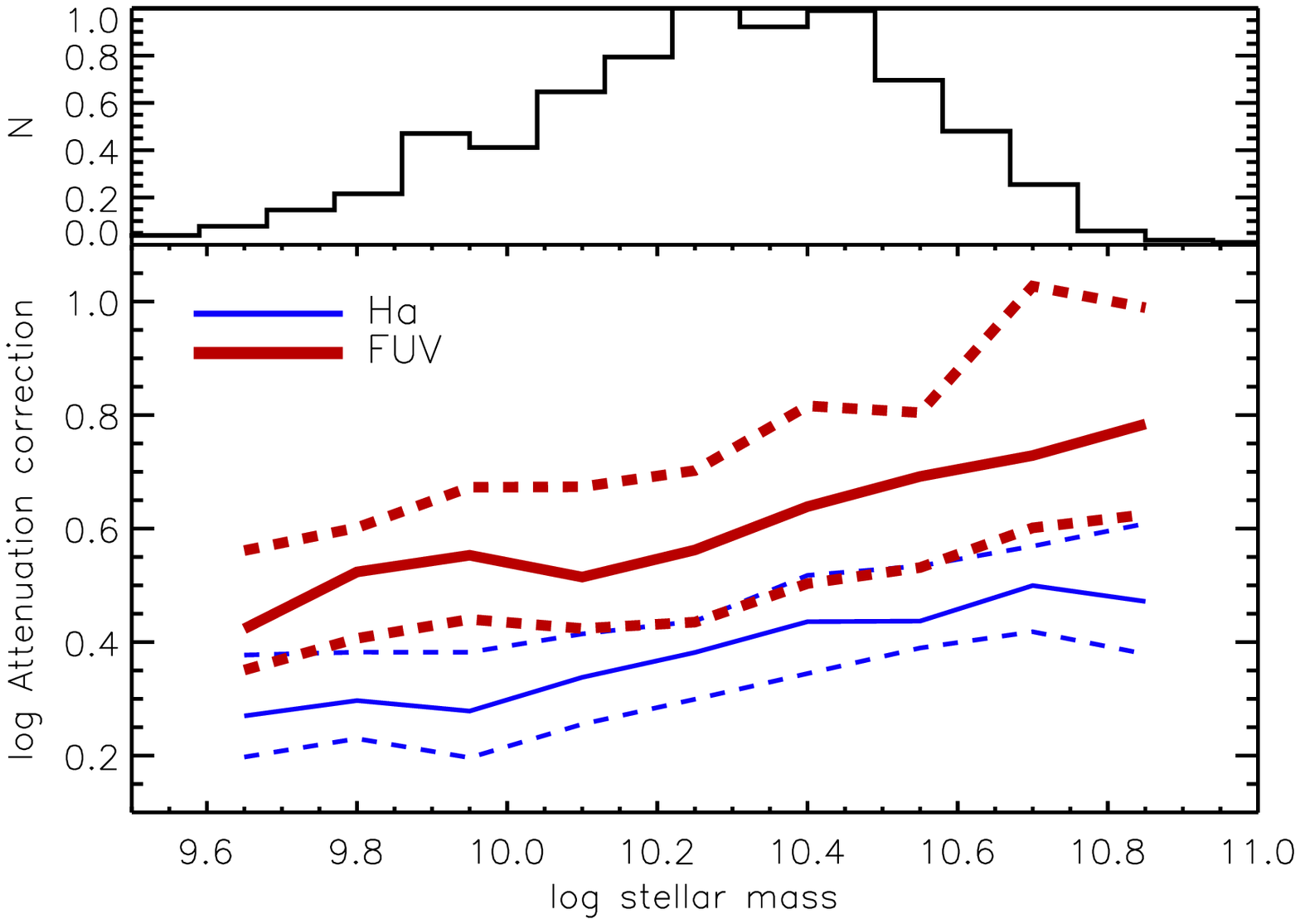}
\includegraphics[height=2.2in,width=3.4in]{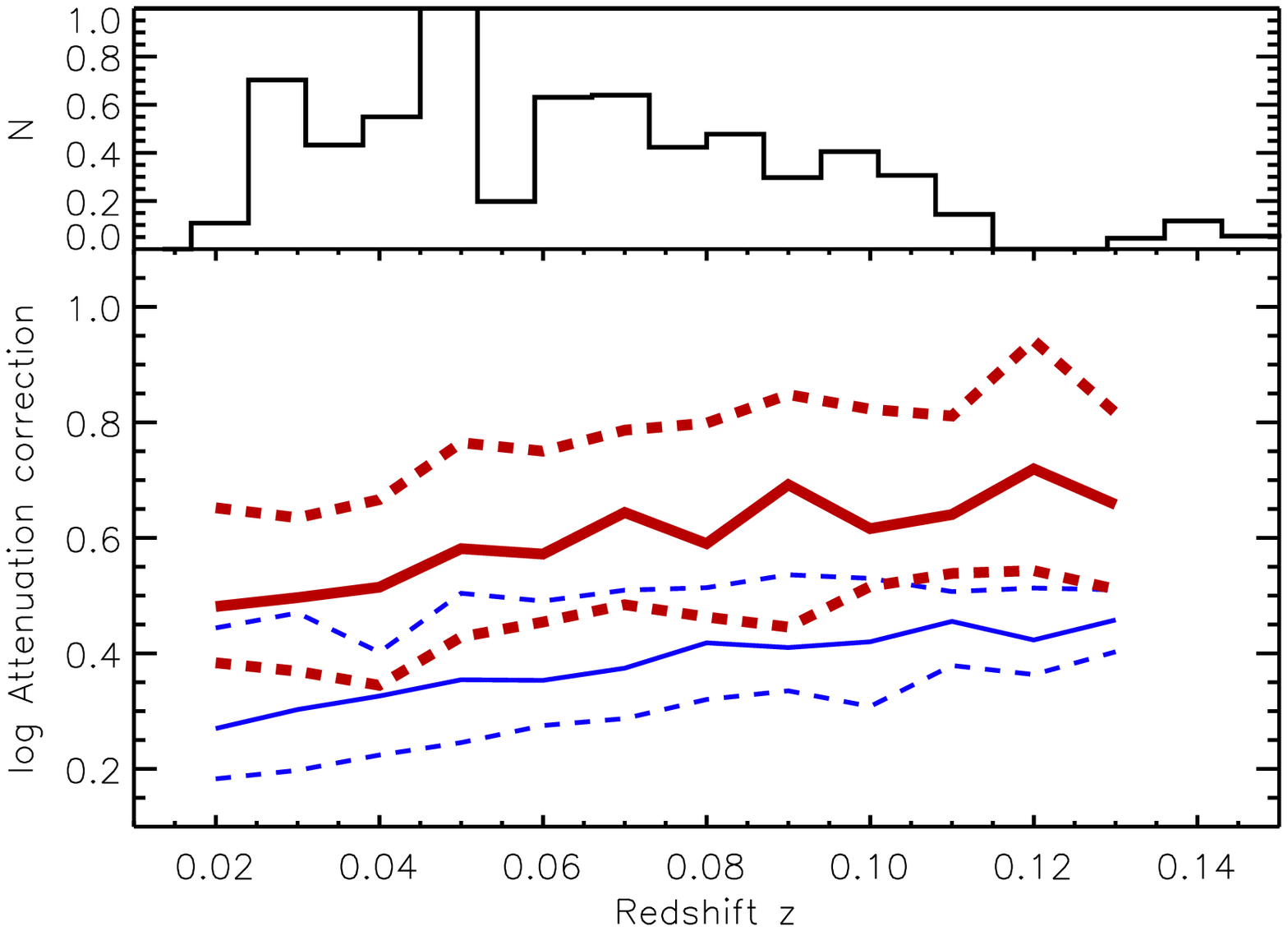}
\includegraphics[height=2.2in,width=3.4in]{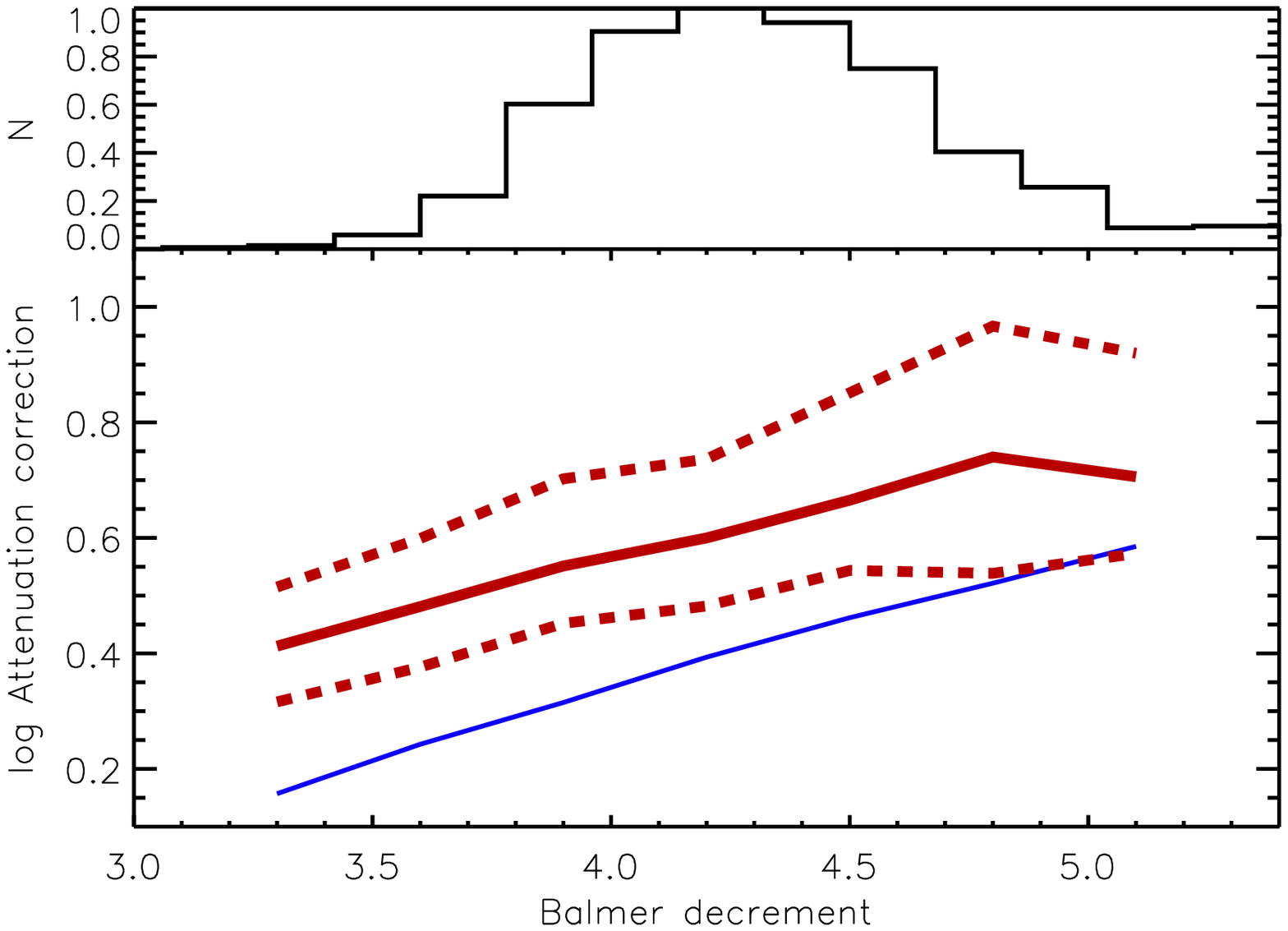}
\includegraphics[height=2.2in,width=3.4in]{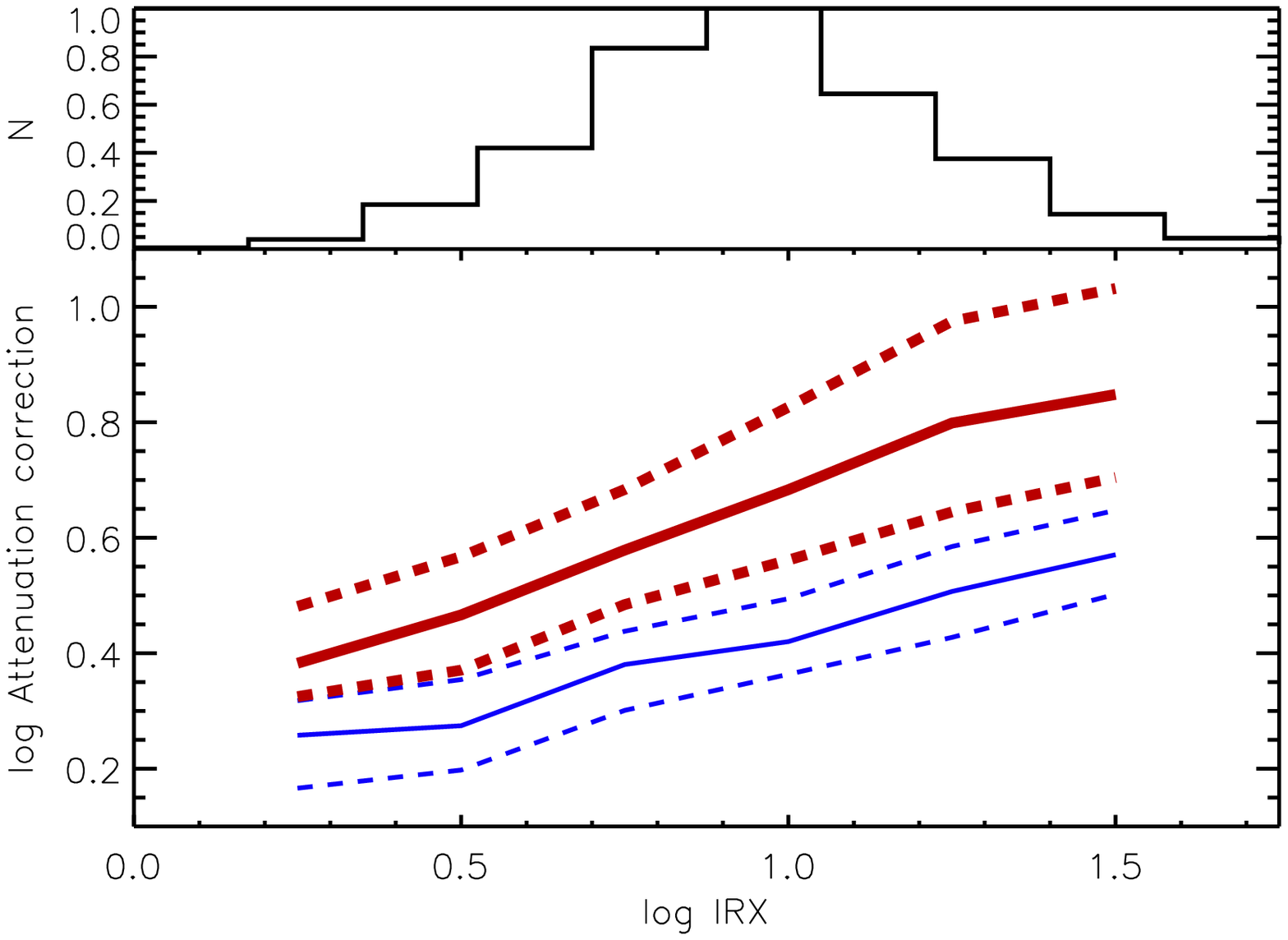}
\includegraphics[height=2.2in,width=3.4in]{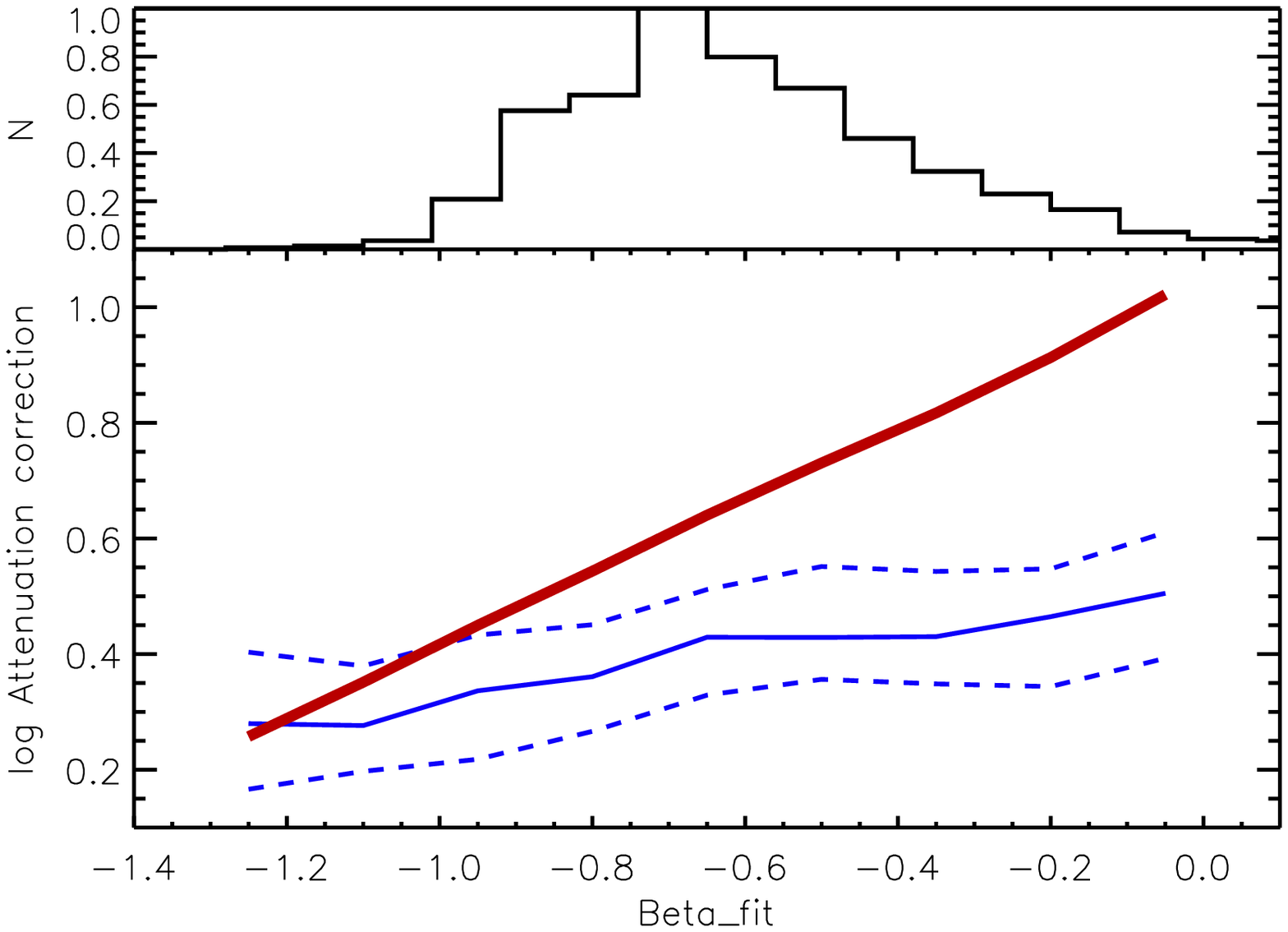}
\includegraphics[height=2.2in,width=3.34in]{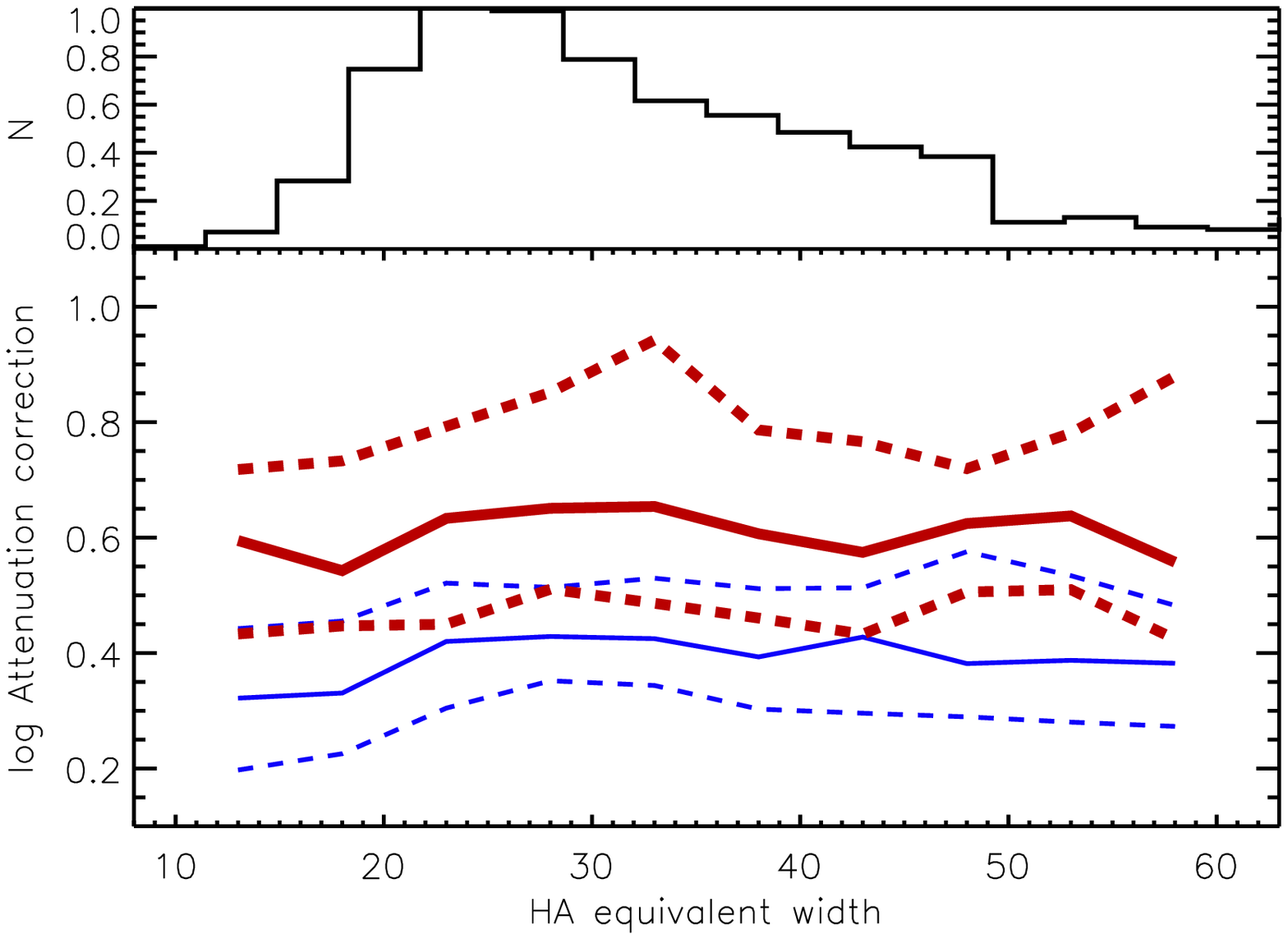}
\includegraphics[height=2.2in,width=3.34in]{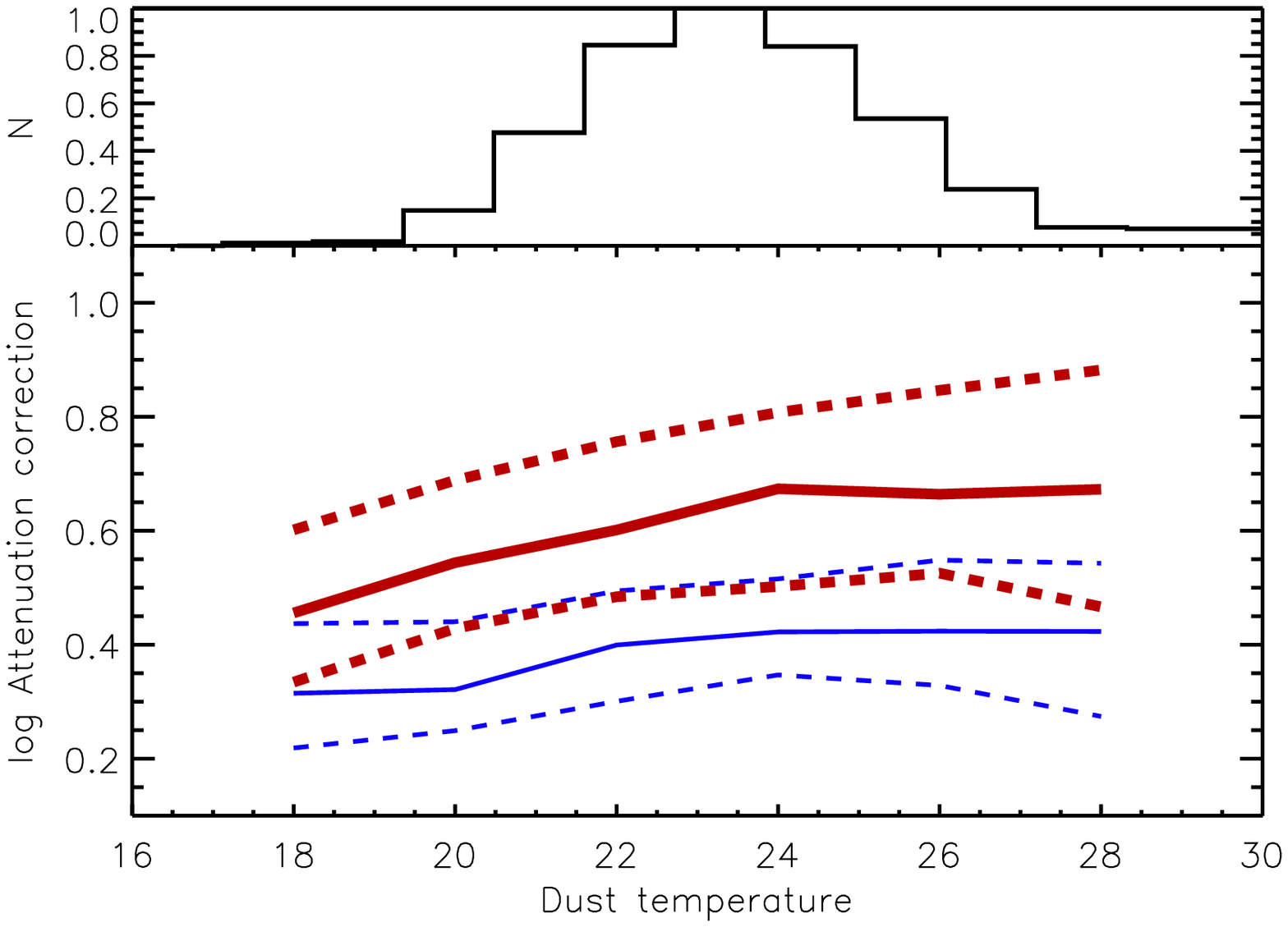}
\includegraphics[height=2.2in,width=3.34in]{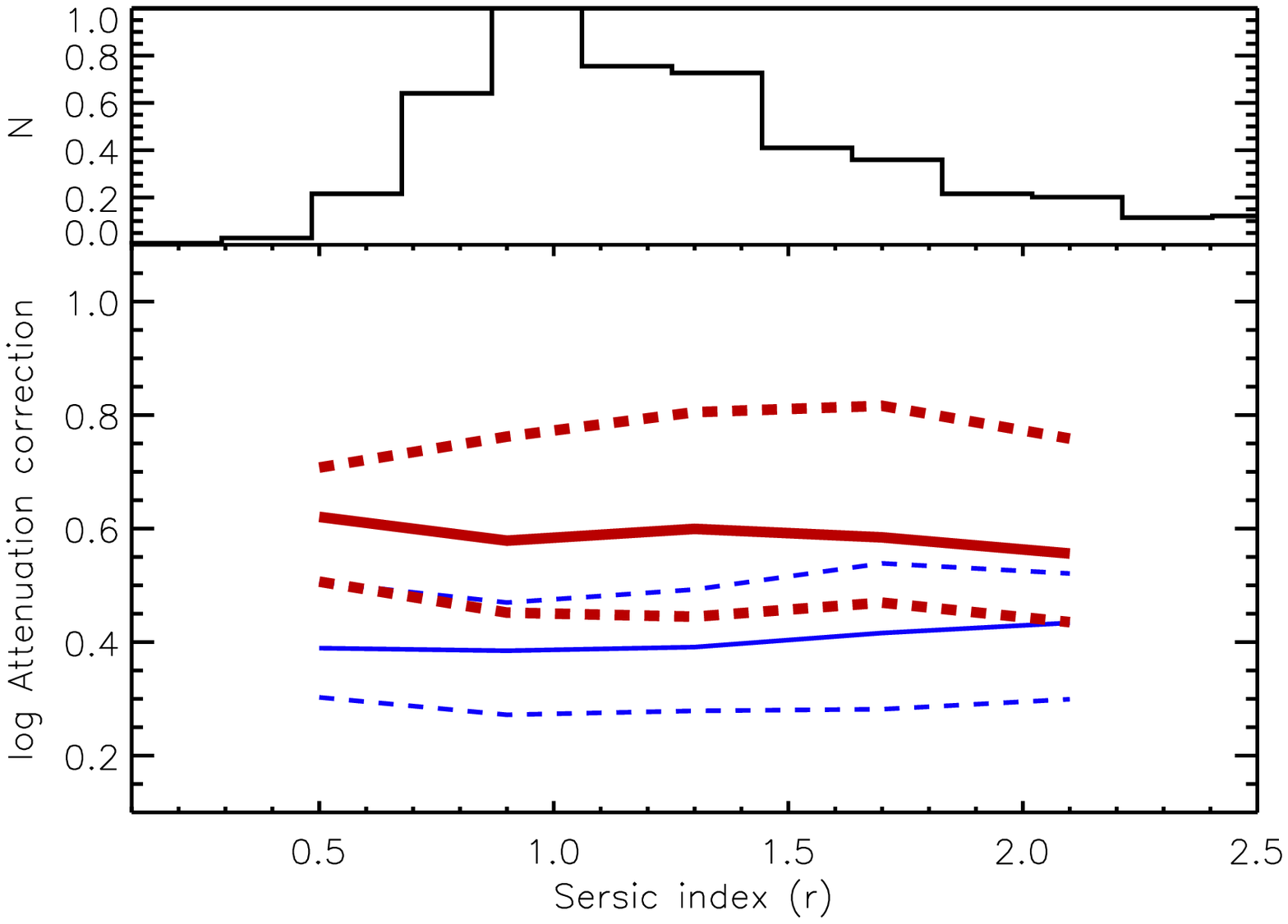}
\caption{The dust attenuation correction factors as a function of various physical parameters (stellar mass, redshift, Balmer decrement, IRX, $\beta_{\rm fit}$, H$\alpha$ equivalent width,  dust temperature, and Sersic index in the r-band). In the top panel of each figure, we show the normalised histogram of the x-axis. The thin blue lines correspond to the 25th, 50th and 75th percentile in the dust correction factors applied in the H$\alpha$-based SFR. The thick red lines correspond to the 25th, 50th and 75th percentile in the  dust correction factors applied in the FUV-based SFR.}
\label{fig:dust_3rdvar}
\end{figure*}

In Fig.~\ref{fig:sfr_comp_3rdvar}, we compare the ratios of different SFR indicators as a function of various physical parameters such as stellar mass, redshift, Balmer decrement, IRX, $\beta_{\rm fit}$, H$\alpha$ equivalent width, dust temperature and S\'ersic index in the SDSS optical bands. In the top panel of each figure, we show the normalised histogram of the x-axis. In the bottom panel of each figure, the blue lines correspond to the 25th, 50th and 75th percentile in the ratio of SFR$_{\rm H\alpha, corr}$ over SFR$_{\rm FUV + IR}$. The red lines correspond to the 25th, 50th and 75th percentile in the ratio of SFR$_{\rm FUV, corr}$ (using $\beta_{\rm fit}$ and our new $A_{\rm FUV}$-$\beta$ relation) over SFR$_{\rm FUV + IR}$.  In Fig.~\ref{fig:dust_3rdvar}, we show the dust attenuation correction factors applied in SFR$_{\rm H\alpha, corr}$ and SFR$_{\rm FUV, corr}$ as a function of these physical parameters.

Stellar mass: Neither SFR$_{\rm FUV, corr}$/SFR$_{\rm FUV + IR}$ nor SFR$_{\rm H\alpha, corr}$/SFR$_{\rm FUV+IR}$ has a significant dependence on stellar mass. The dust attenuation corrections applied in SFR$_{\rm FUV, corr}$ and SFR$_{\rm H\alpha, corr}$ increase with increasing stellar mass indicating more massive galaxies are more obscured. It could also imply that galaxies at higher redshift are more obscured as more massive galaxies are preferentially located at higher redshift in our sample (selected from flux-limited surveys). We need a larger sample to properly disentangle the effect of stellar mass and redshift.

Redshift: Neither SFR$_{\rm FUV, corr}$/SFR$_{\rm FUV + IR}$ nor SFR$_{\rm H\alpha, corr}$/SFR$_{\rm FUV+IR}$ exhibits a significant dependence on redshift. The attenuation corrections applied in SFR$_{\rm FUV, corr}$ and SFR$_{\rm H\alpha, corr}$ increase with increasing redshift. It could be because galaxies at higher redshift are more obscured and/or we are preferentially selecting more massive galaxies at higher redshift.

Balmer decrement (BD): The attenuation correction in SFR$_{\rm H\alpha, corr}$ is uniquely determined by BD. The attenuation correction applied in SFR$_{\rm FUV, corr}$ increases with increasing BD which is expected given the correlation between A(H$\alpha$) and $\beta$ seen in the top panel in Fig. 7.  SFR$_{\rm H\alpha, corr}$/SFR$_{\rm FUV+IR}$ increases with increasing values of BD (i.e. larger attenuation correction) which is most likely due to the fact that the dust-corrected H$\alpha$ line luminosity directly depends on BD. The SFR$_{\rm FUV, corr}$/SFR$_{\rm FUV + IR}$ ratio decreases with increasing BD, which is caused by the broad correlation between BD and IRX and the dependence of SFR$_{\rm FUV, corr}$/SFR$_{\rm FUV + IR}$ on IRX (see below).

IRX: IRX is a measure of the overall UV photon escape fraction. As expected, the attenuation corrections applied in SFR$_{\rm FUV, corr}$ and SFR$_{\rm H\alpha, corr}$ increase with increasing IRX. This is consistent with Fig. 7. SFR$_{\rm H\alpha, corr}$/SFR$_{\rm FUV + IR}$ shows no appreciable dependence on IRX. SFR$_{\rm FUV, corr}$/SFR$_{\rm FUV + IR}$ decreases significantly with increasing IRX, which is caused by the large scatter in the IRX - $\beta$ correlation shown in the bottom panel in Fig. 7. For objects with high IRX values, it is clear that the dust correction factors $A_{\rm FUV}$ based on the IRX - $\beta$ correlation derived for the whole sample (the red line in the bottom panel in Fig. 7) will underestimate the true level of attenuation.

UV continuum slope $\beta_{\rm fit}$: The attenuation correction applied in SFR$_{\rm FUV, corr}$ is uniquely determined by $\beta$. The attenuation correction applied in SFR$_{\rm H\alpha, corr}$ increases with increasing $\beta$ which is expected given the correlation between A(H$\alpha$) and $\beta$. Both SFR$_{\rm H\alpha, corr}$/SFR$_{\rm FUV+IR}$ and SFR$_{\rm FUV, corr}$/SFR$_{\rm FUV + IR}$ stays more or less flat with changing $\beta_{\rm fit}$.

H$\alpha$ equivalent width (EW): The H$\alpha$ emission line EW is a measure of specific SFR (Kenicutt et al. 1994).  The attenuation correction factors applied in SFR$_{\rm H\alpha, corr}$ and SFR$_{\rm FUV, corr}$ do not vary as a function of the H$\alpha$ EW. SFR$_{\rm FUV, corr}$/SFR$_{\rm FUV+IR}$ stays flat. SFR$_{\rm H\alpha, corr}$/SFR$_{\rm FUV + IR}$ increases with increasing H$\alpha$ EW. One explanation could be that measurement of the H$\alpha$ EW affects the H$\alpha$ line luminosity.

Dust temperature: The attenuation corrections applied in SFR$_{\rm FUV, corr}$ and SFR$_{\rm H\alpha, corr}$ increase slightly with increasing dust temperature. This is consistent with the fact that galaxies with warmer dust tend to have higher infrared luminosities. The SFR$_{\rm H\alpha, corr}$ /SFR$_{\rm FUV+IR}$ ratio does not have a significant dependence on dust temperature. However, the SFR$_{\rm FUV, corr}$/SFR$_{\rm FUV + IR}$ ratio decreases with increasing dust temperature which is due to the broad positive correlation between dust temperature and IRX. As explained above, SFR$_{\rm FUV, corr}$/SFR$_{\rm FUV + IR}$ decreases with increasing IRX caused by the large scatter in the IRX - $\beta$ correlation.

S\'ersic index: Neither the attenuation correction factors or the SFR ratios changes significantly a a function of the Sersic index in the r-band. We find similar trends with respect to the Sersic indexes in the other SDSS optical bands. 

\section{Summary}

In this paper, we compare multi-wavelength star-formation rate (SFR) indicators in the local Universe in the three GAMA equatorial fields. Our analysis uses ultraviolet (UV) photometry from GALEX,  far-infrared (FIR) and sub-millimetre (sub-mm) photometry from {\it Herschel} H-ATLAS, and H$\alpha$ spectroscopy from the GAMA redshift survey. To minimise random statistical errors, we construct a very high quality sample of 745 objects (median redshift $\left<z\right>=0.08$). We consider three commonly used SFR indicators: UV continuum luminosity corrected for dust attenuation using the UV spectral slope (SFR$_{\rm UV, corr}$), H$\alpha$ emission line luminosity corrected for dust attenuation using the Balmer decrement (SFR$_{\rm H\alpha, corr}$), and the combination of UV and infrared dust emission (SFR$_{\rm UV + IR}$).  

We find a good linear correlation between SFR$_{\rm UV, corr}$ and SFR$_{\rm UV + IR}$ but with a $\sim0.3$ dex offset when using the UV spectral slope $\beta$ and the Hao et al. (2011) $A_{\rm FUV}$-$\beta$ relation for deriving the dust attenuation correction. This offset is removed when we replace the Hao et al. relation with our new $A_{\rm FUV}$-$\beta$  relation based on calibrating IRX and the attenuation in H$\alpha$ against $\beta$. The $A_{\rm FUV}$-$\beta$ relation is slightly different depending on whether $\beta_{\rm fit}$ or $\beta_{\rm colour}$ is used and the choice of IR SED library. The difference between the Hao et al. (2011) $A_{\rm FUV}$-$\beta$ relation based on a nearby star-forming sample and the new relation derived in this paper is due to the difference in the galaxy samples. In addition to being at higher redshifts, our galaxy sample corresponds to much lower survey flux limits in the IR and UV and therefore contains many more quiescent star-forming galaxies with redder UV spectral slopes and lower IRX values. We also find a good linear correlation between SFR$_{\rm H\alpha, corr}$ and SFR$_{\rm UV + IR}$. There is a small median offset of around 0.1 dex. But we demonstrate that this offset can be entirely explained by systematic effects in deriving the infrared luminosity $L_{\rm IR}$ and/or other systematic errors in the H$\alpha$-based SFR tracer. Moreover, the correlation between SFR$_{\rm H\alpha, corr}$ and SFR$_{\rm UV + IR}$ has a similar scatter (0.2 dex) as the correlation between SFR$_{\rm UV, corr}$ and SFR$_{\rm UV + IR}$.

The ratios between different SFR indicators and the dust attenuation correction factors applied in the UV (using $\beta$) and H$\alpha$ (using the Balmer decrement) are examined as a function of various galaxy physical parameters. The attenuation factor applied in SFR$_{\rm H\alpha, corr}$ which is uniquely determined by Balmer decrement increases with increasing values of IRX and $\beta$. Similarly, the attenuation factor applied in SFR$_{\rm UV, \beta}$ which is uniquely determined by $\beta$ increases with increasing values of Balmer decrement and IRX. These trends are consistent with the broad correlations between Balmer decrement, $\beta$, and IRX seen in Fig. 7. We also find that attenuation correction factors depends on stellar mass, redshift and dust temperature, but not on the H$\alpha$ equivalent width or Sersic index in the SDSS optical bands.

After applying corrections for dust attenuation, we find that the SFR$_{\rm UV, corr}$/SFR$_{\rm UV + IR}$ ratio does not depend significantly on stellar mass, redshift,  UV spectral slope $\beta$, H$\alpha$ equivalent width, or structural parameters such as S\'ersic index. However, the SFR$_{\rm UV, corr}$/SFR$_{\rm UV + IR}$ ratio does systematically decrease with increasing values of IRX, Balmer decrement, and dust temperature. The dependence on IRX is caused by the large scatter in the IRX vs $\beta$ relation. For objects with high IRX values, the dust attenuation correction factor $A_{\rm FUV}$ based on the IRX - $\beta$ correlation derived for the whole sample will underestimate the true level of attenuation. Also, there is a positive correlation between IRX and Balmer decrement and between IRX and dust temperature which explains the systematic trend in the SFR$_{\rm UV, corr}$/SFR$_{\rm UV + IR}$ ratio as a function of Balmer decrement and dust temperature. In contrast, the SFR$_{\rm H\alpha, corr}$/SFR$_{\rm UV + IR}$ ratio does not show any systematic trend as a function of various physical parameters except Balmer decrement and H$\alpha$ equivalent width, which is most likely caused by the fact that both Balmer decrement and H$\alpha$ equivalent width directly determine the dust-corrected H$\alpha$ line luminosity.

\section*{ACKNOWLEDGEMENTS}

We thank the anonymous referee whose comments improved the paper significantly. LW acknowledges support from an ERC StG grant (DEGAS-259586). PN acknowledges the support of the Royal Society through the award of a University Research Fellowship, the European Research Council, through receipt of a Starting Grant (DEGAS-259586) and support of the Science and Technology Facilities Council (ST/L00075X/1). SB acknowledges the funding support from the Australian Research Council  through a Future Fellowship (FT140101166). NB acknowledges the support of the EC FP7 SPACE project ASTRODEEP (Ref. No. 312725). RJI, LD, SJM and IO acknowledge support from ERC in the form of the Advanced Investigator Programme, 321302, COSMICISM. E. Ibar acknowledges funding from CONICYT/FONDECYT postdoctoral project N$^\circ$:3130504. MALL acknowledges support from UNAM through the PAPIIT project IA101315. SD acknowledges support from a STFC Ernest Rutherford fellowship.

GAMA is a joint European-Australasian project based around a spectroscopic campaign using the Anglo-Australian Telescope. The GAMA input catalogue is based on data taken from the Sloan Digital Sky Survey and the UKIRT Infrared Deep Sky Survey. Complementary imaging of the GAMA regions is being obtained by a number of independent survey programs including GALEX MIS, VST KiDS, VISTA VIKING, WISE, Herschel-ATLAS, GMRT and ASKAP providing UV to radio coverage. GAMA is funded by the STFC (UK), the ARC (Australia), the AAO, and the participating institutions. The GAMA website is http://www.gama-survey.org/.

The H-ATLAS is a project with Herschel, which is an ESA space observatory with science instruments provided by European-led Principal Investigator consortia and with important participation from NASA. The H-ATLAS web site is \url{http://www.h-atlas.org/}.

We acknowledge the use of data products from the NASA operated GALEX space mission. We also acknowledge the following institutions and agencies for their financial contributions towards the reactivation and operations of the GALEX satellite. This has allowed us to complete NUV observations of the G23 and Herschel ATLAS SGP regions: The Australian-Astronomical Observatory (AAO), the Australian Research Council (ARC), the International Centre for Radio Astronomy Research (ICRAR), the University of Western Australia, the University of Sydney, the University of Canterbury, Max Plank Institute fuer Kernphysik (MPIK), the University of Queensland, the University of Edinburgh, Durham University, the European Southern Observatory (ESO), the University of Central Lancashire, Liverpool John Moore University, National Aeronautics and Space Administration (NASA), Universit\'{e} Paris Sud, University of California Irvine, Instituto Nazionale Di Astrofisica (INAF), and the University of Hertfordshire.

\appendix

\section{H-ATLAS blind catalogue}
\label{appendix1}

\begin{figure*}
\includegraphics[height=2.4in,width=3.3in]{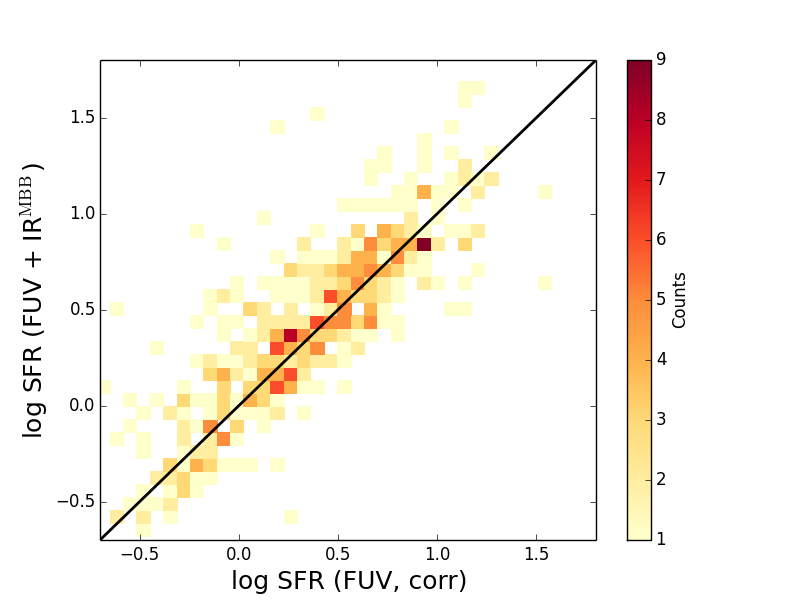}
\includegraphics[height=2.4in,width=3.3in]{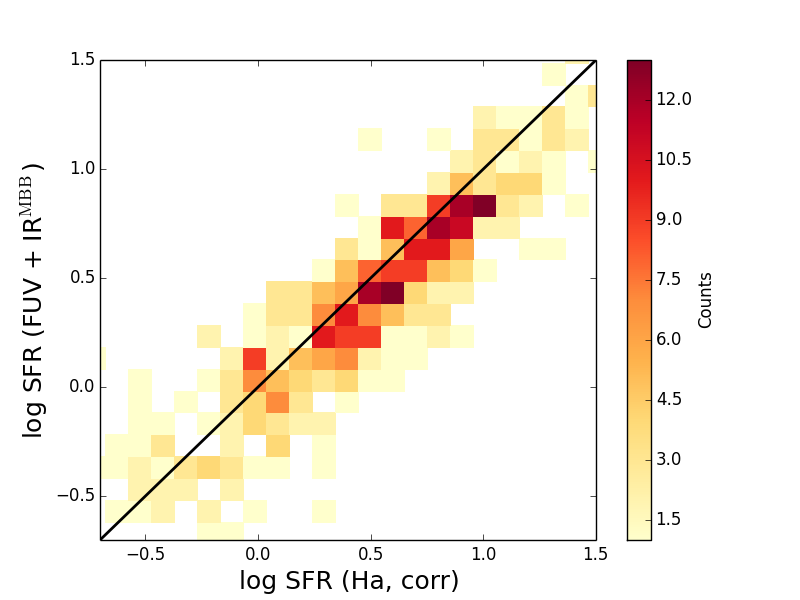}
\caption{Comparison of $\log_{10} {\rm SFR}_{\rm FUV, corr}$ (using $\beta_{\rm fit}$) with $\log_{10} {\rm SFR}_{\rm FUV + IR}$ (left panel) and $\log_{10} {\rm SFR}_{\rm H\alpha, corr}$ with $\log_{10} {\rm SFR}_{\rm FUV + IR}$ (right panel) (colour-coded by galaxy counts). The H-ATLAS catalogue is matched to the GAMA galaxies using the likelihood ratio (LR) technique. The black line is the one-to-one relation.}
\label{fig:SFR_comp_uv_LR}
\end{figure*}

The large PSF of {\it Herschel} imaging and the relatively poor correlation between sub-mm and optical brightness means that matching sub-mm sources to optical sources can be ambiguous. The two techniques considered represent alternative solutions to the problem: the technique used in the main body of the paper is forced sub-mm photometry at known optical source positions, while the alternative method is likelihood-ratio matching between sources extracted independently from sub-mm and optical images. The first technique can be prone to bias when bright sub-mm sources are missing from the optical prior, while the second is biased against sub-mm undetected sources as well as those with ambiguous optical counterparts. Therefore, it is important to test whether the correlations we see between SFR$_{\rm UV+IR}$ and SFR$_{\rm UV, corr}$ and between SFR$_{\rm UV+IR}$, and  SFR$_{\rm H\alpha, corr}$ depend on how {\it Herschel} fluxes are derived for GAMA galaxies. Here we re-derive the correlations between different SFR indicators using the H-ATLAS blind source catalogue.

The H-ATLAS catalogue contains all SPIRE sources which are $>5\sigma$ significance (including confusion noise) in any of the three SPIRE bands (250, 350, 500 $\mu$m). PACS fluxes (100, 160 $\mu$m) are measured using circular apertures placed at the SPIRE positions. The source catalogue is based on finding peaks in the noise-weighted PSF filtered the maps using the MADX algorithm. Please refer to Rigby et al. (2011), Maddox et al. (in prep.) and Valiante et al. (in prep.) for details of the source extraction method. The blind H-ATLAS catalogue is matched to GAMA galaxies using the likelihood ratio (LR) method (e.g., Sutherland \& Saunders 1992; Chapin et al. 2011; Wang \& Rowan-Robinson 2009, 2014). Briefly, the LR method uses the positional and brightness information to identify the most likely GAMA counterpart to an H-ATLAS source. For more details, we refer the reader to Smith et al. (2011) and Bourne et al. (in prep.).

In Fig. ~\ref{fig:SFR_comp_uv_LR}, we show the correlations between different SFR indicators, using the H-ATLAS blind source catalogue matched with GAMA galaxies through the LR method. The resulting correlations are very similar to Fig.~\ref{fig:new_SFR_comp_uv} and Fig.~\ref{fig:SFR_comp_ha} in Section 4.3.  Note that we have applied our new $A_{\rm FUV}$-$\beta_{\rm fit}$ relation (i.e. Eq 22) and not the Hao et al. (2011) relation.

\section{Comparison with the Hao et al. (2012) sample and relation}
\label{appendix2}

To understand the differences seen in the IRX vs $\beta$ relation between Hao et al. (2011) and this paper, we compare in detail the galaxy samples used in both studies. 

\begin{figure}
\includegraphics[height=2.4in,width=3.4in]{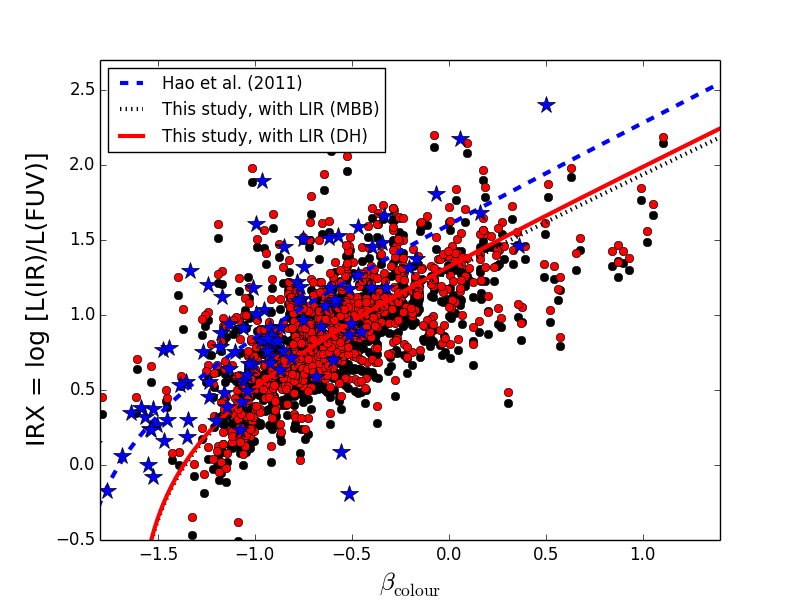}
\caption{IRX (i.e. $\log_{10} L_{\rm IR}/L_{\rm FUV}$) vs $\beta_{\rm colour}$. To better compare Hao et al. (2011) and this paper, $\beta_{\rm colour}$ is used in this figure. The blue stars correspond to the MK06 sample used in Hao et al. (2011) and the blue dashed line is the best-fit from Hao et al. (2011).  The black symbols correspond to our joint UV-H$\alpha$-IR sample of 745 objects using IR luminosities measured from SED fitting to the MBB library. The black dotted line is our best-fit IRX-$\beta$ relation to the black symbols. To better match the IR luminosity measurement from Hao et al. (2012), we also plot our measurement using IR luminosities estimated from SED fitting to the Dale \& Helou (2002) library which are shown as the red symbols. The red solid line is the best fit to the red symbols.}
\label{fig:dust_atten_mk}
\end{figure}

Fig.~\ref{fig:dust_atten_mk} shows IRX vs $\beta$ using the Moustakas \& Kennicutt (2006; hereafter MK06) galaxy sample studied in Hao et al. (2011) and our joint UV-H$\alpha$-IR sample of 745 objects. To better compare the Hao et al. (2011) study and this paper, $\beta_{\rm colour}$ is used in this figure instead of $\beta_{\rm fit}$. The blue stars correspond to the MK06 sample and the blue line is the best-fit of the functional form defined in Eq. 19 from Hao et al. (2011) with $s_{\rm FUV=3.83}$ and $a_{\rm FUV=0.46}$.  The black symbols correspond to our joint UV-H$\alpha$-IR sample of 745 objects using IR luminosities measured from SED fitting to the MBB library. The black line is our best-fit IRX-$\beta$ relation to the black symbols with $s_{\rm FUV=3.55}$ and $a_{\rm FUV=0.46}$. To better match the IR luminosity measurement from Hao et al. (2012), we also plot our measurement using IR luminosities estimated from SED fitting to the Dale \& Helou (2002) library which are shown as the red symbols. The red line is the best fit to the red symbols with $s_{\rm FUV=3.66}$ and $a_{\rm FUV=0.46}$. As discussed in Section 3.3, $L_{\rm IR}$ estimated using the DH library is systematically higher than $L_{\rm IR}$ estimated from the MBB library. However, the median difference is less than 0.1 dex between the two libraries. Comparing the blue stars with the red symbols, it is clear that our joint UV-H$\alpha$-IR sample has a lot more quiescent star-forming galaxies with redder UV spectra and lower IRX values.

\begin{figure}
\includegraphics[height=2.4in,width=3.4in]{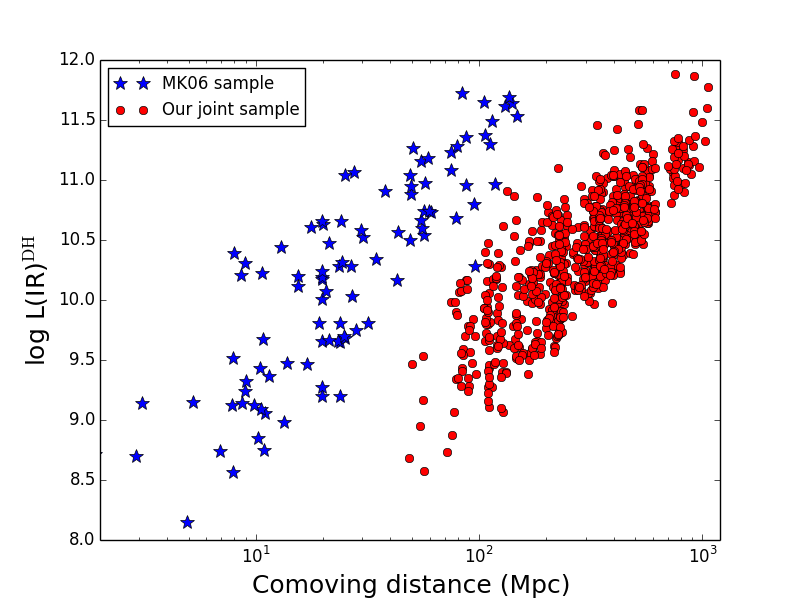}
\includegraphics[height=2.4in,width=3.4in]{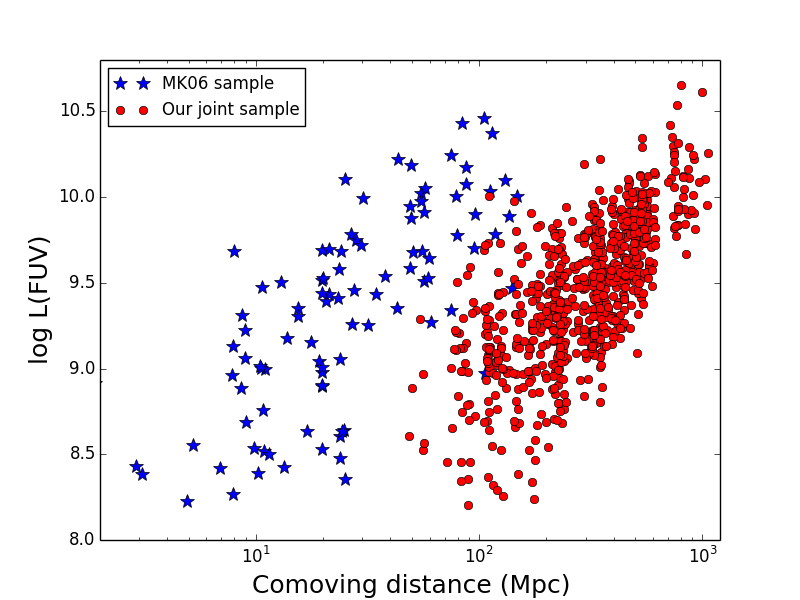}
\caption{Top: Infrared luminosity $L_{\rm IR}$ (in unit of $L_{\odot}$) vs comoving distance. The blue stars correspond to the MK06 galaxy sample. The red symbols correspond to our joint UV-H$\alpha$-IR sample with $L_{\rm IR}$ estimated from SED fitting to the  Dale \& Helou (2002) library. Bottom: Similar to the top panel but with the y-axis replaced by the observed FUV luminosity $L_{\rm FUV}$ uncorrected for dust attenuation.}
\label{fig:lir_mk}
\end{figure}

In Fig.~\ref{fig:lir_mk}, we compare bolometric $L_{\rm IR}$ vs coming distance and the observed FUV luminosity $L{\rm FUV}$ (without correction for dust) vs coming distance between the MK06 sample and our joint UV-H$\alpha$-IR sample. The galaxies in our sample are at much higher redshifts than the MK06 sample. In addition, our galaxies correspond to  a much lower survey flux limit than the MK06 sample. This is mostly like due to the fact that the MK06 sample are restricted to galaxies which are detected by IRAS at 25, 60 and 100 $\mu$m. As such, the MK06 sample is biased towards warmer dust temperature and more infrared luminous galaxies than our galaxy sample.

\section{Comparison with SFR derived from radiative transfer modelling}
\label{appendix3}

\begin{figure}
\includegraphics[height=2.3in,width=3.4in]{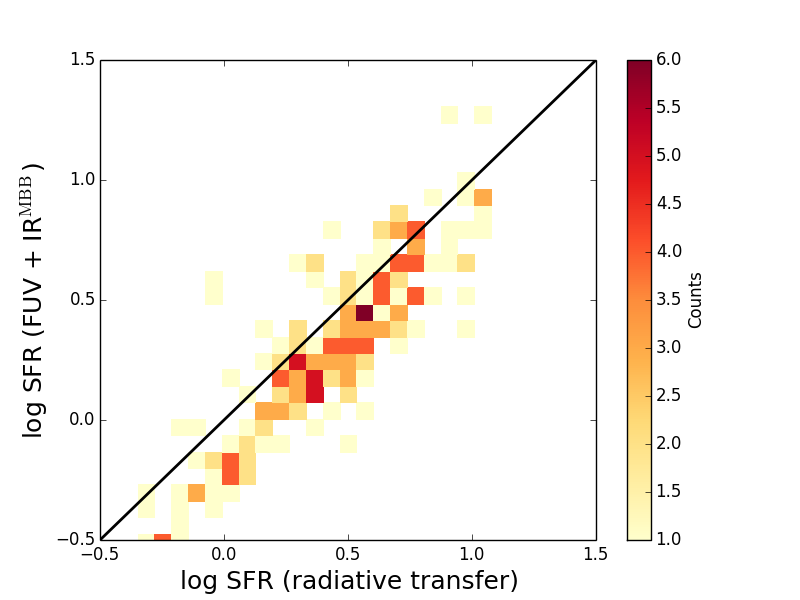}
\includegraphics[height=2.3in,width=3.4in]{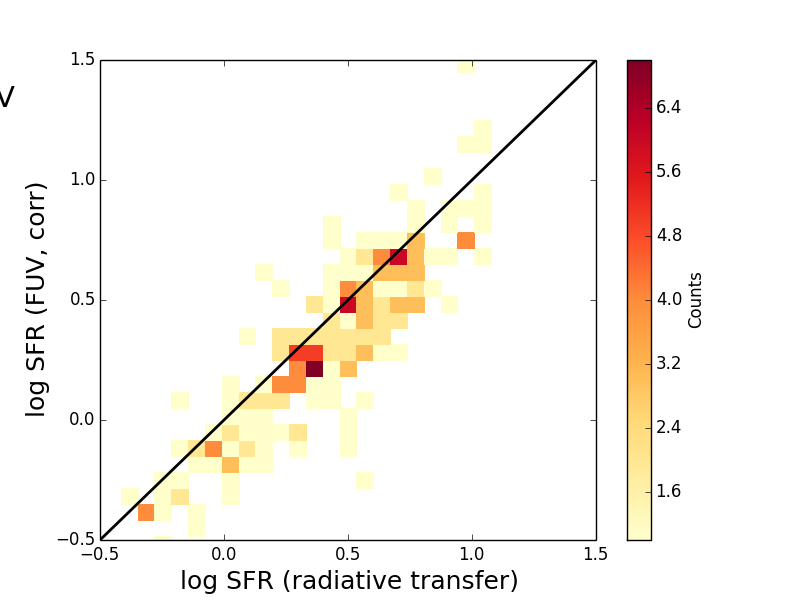}
\includegraphics[height=2.3in,width=3.4in]{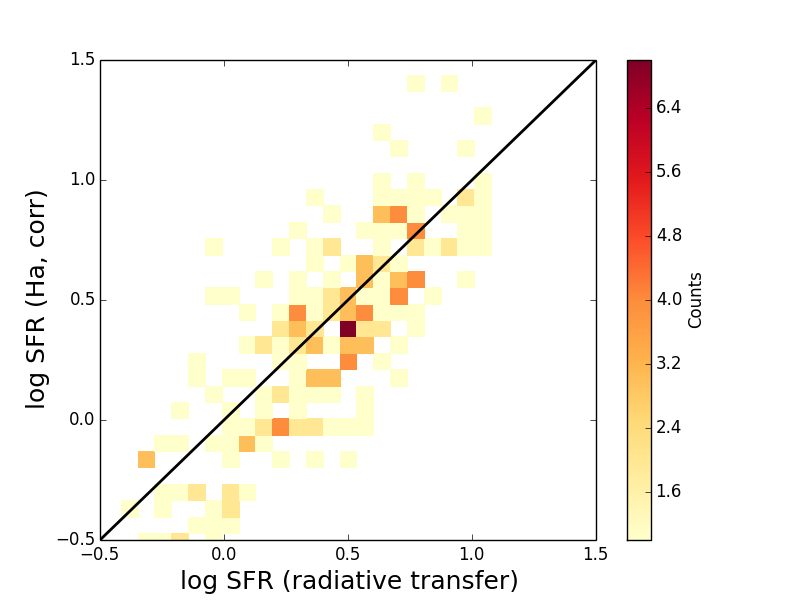}
\caption{Top: Comparison of NUV-based ${\rm SFR}$ (in unit of $M_{\odot}$/yr) derived from radiative transfer modelling with ${\rm SFR}_{\rm FUV + IR^{\rm MBB}}$ (colour-coded by galaxy counts). Middle: Comparison of NUV-based ${\rm SFR}$ derived from radiative transfer modelling with ${\rm SFR}_{\rm FUV, corr}$. Bottom: Comparison of NUV-based ${\rm SFR}$ derived from radiative transfer modelling with ${\rm SFR}_{\rm H\alpha, corr}$. The black line in all panels is the one-to-one relation.}
\label{fig:SFR_comp_radiative}
\end{figure}

Arguably an accurate determination of SFR requires radiative transfer modelling of the panchromatic SEDs of galaxies, which could then be used to calibrate SFRs derived from other SFRs indicators. In Fig.~\ref{fig:SFR_comp_radiative}, we compare our SFR indicators (SFR$_{\rm UV + IR^{\rm MBB}}$, SFR$_{\rm UV, corr}$ and SFR$_{\rm H\alpha, corr}$) with the NUV-based SFRs (Grootes et al. 2013) derived using the radiative transfer models of Popescu et al. (2011) for a sample of local GAMA spiral galaxies at $z<0.13$, SFR$_{\rm RT}$. The black line indicates the one-to-one relation. In Table B1, we  list the 16th, 50th and 84th percentile in the difference between the SFR indicators (${\rm SFR_{UV+IR^{\rm MBB}}}$,   ${\rm SFR_{UV, corr}}$ and ${\rm SFR_{H\alpha, corr}}$) studied in this paper and the the radiation transfer corrected NUV based SFR$_{\rm RT}$. There is a small median difference in all cases except in ${\rm SFR_{NUV, corr}}$ - SFR$_{\rm RT}$, which could be because the radiation transfer corrected SFR is also derived from the observed NUV luminosity. Finally, the correlation between SFR$_{\rm H\alpha, corr}$  and SFR$_{\rm RT}$  has considerably larger scatter compared to the correlations seen between SFR$_{\rm UV + IR^{\rm MBB}}$ and SFR$_{\rm RT}$ and between SFR$_{\rm UV, corr}$ and SFR$_{\rm RT}$.

\begin{table}
\caption{The 16th, 50th and 84th percentile in the difference between the SFR indicators (${\rm SFR_{UV+IR}}$,   ${\rm SFR_{UV, corr}}$ and ${\rm SFR_{H\alpha, corr}}$) studied in this paper and the NUV based SFRs (Grootes et al. 2013) derived using the radiative transfer (RT) models, ${\rm SFR_{RT}}$.}\label{table:beta}
\begin{tabular}{ll}
\hline
SFR$_{\rm FUV+IR^{\rm MBB}}$ - ${\rm SFR_{RT}}$ & -0.3, -0.2, 0.0\\
\hline 
SFR$_{\rm NUV+IR^{\rm MBB}}$ - ${\rm SFR_{RT}}$ & -0.2, -0.1, 0.0\\
\hline
SFR$_{\rm FUV, corr}$ - ${\rm SFR_{RT}}$ &  -0.3, -0.1, 0.0\\
\hline
SFR$_{\rm NUV, corr}$ - ${\rm SFR_{RT}}$ & -0.1, 0.0, 0.2\\
\hline 
SFR$_{\rm H\alpha, corr}$ - ${\rm SFR_{RT}}$ & -0.3, -0.1, 0.2\\
\hline
\end{tabular}
\end{table}

\end{document}